\documentclass[pmlr,twocolumn,10pt]{jmlr}

\usepackage{microtype}
\usepackage{graphicx}
\usepackage{booktabs} 
\usepackage[switch]{lineno}

\usepackage{hyperref}
\usepackage{balance}

\usepackage{amsmath}
\usepackage{amsfonts}
\usepackage{bbm}
\usepackage{multirow}
\usepackage{enumitem}
\usepackage{xcolor}

\newcommand{\indep}{\perp \!\!\! \perp}
\newtheorem{asm}{Assumption}
\DeclareMathOperator*{\argmax}{arg\,max}

\jmlrvolume{297}
\jmlryear{2025}
\jmlrworkshop{Machine Learning for Health (ML4H) 2025} 

\title{Identifying treatment response subgroups\\in observational time-to-event data}
\author{
\Name{Vincent Jeanselme} \Email{vj2292@columbia.com}\\
\addr University of Cambridge, The Alan Turing Institute\\
\Name{Chang Ho Yoon}
\and
\Name{Fabian Falck}\\
\addr University of Oxford, The Alan Turing Institute\\
\Name{Brian Tom}\and
\Name{Jessica Barrett}\\
\addr University of Cambridge
}

\begin{document}
\maketitle

\begin{abstract}
Identifying patient subgroups with different treatment responses is an important task to inform medical recommendations, guidelines, and the design of future clinical trials. 
Existing approaches for treatment effect estimation primarily rely on Randomised Controlled Trials (RCTs), which tend to feature more homogeneous patient groups, making them less relevant for uncovering subgroups in the population encountered in real-world clinical practice. 
Subgroup analyses established for RCTs suffer from significant statistical biases when applied to observational studies, which benefit from larger and more representative populations.  
Our work introduces a novel, outcome-guided, subgroup analysis strategy for identifying subgroups of treatment response in both RCTs and observational studies alike. 
It hence positions itself in-between individualised and average treatment effect estimation to uncover patient subgroups with distinct treatment responses, critical for actionable insights that may influence treatment guidelines.
In experiments, our approach significantly outperforms the current state-of-the-art method for subgroup analysis in both randomised and observational treatment regimes.
\end{abstract}


\paragraph*{Data and Code Availability}
Experiments are performed on synthetic data and a publicly available dataset from the Surveillance, Epidemiology, and End Results Program\footnote{Available at \url{https://seer.cancer.gov/}}. The code to reproduce the proposed model and the presented results is available on GitHub\footnote{\url{https://github.com/Jeanselme/CausalNeuralSurvivalClustering}}.

\paragraph*{Institutional Review Board (IRB)}
This research does not require IRB approval as it relies on a publicly available dataset from previously approved studies.

\section{Introduction}
\label{sec:intro}
\begin{figure*}[t!] 
    \centering
    \includegraphics[width=0.75\textwidth]{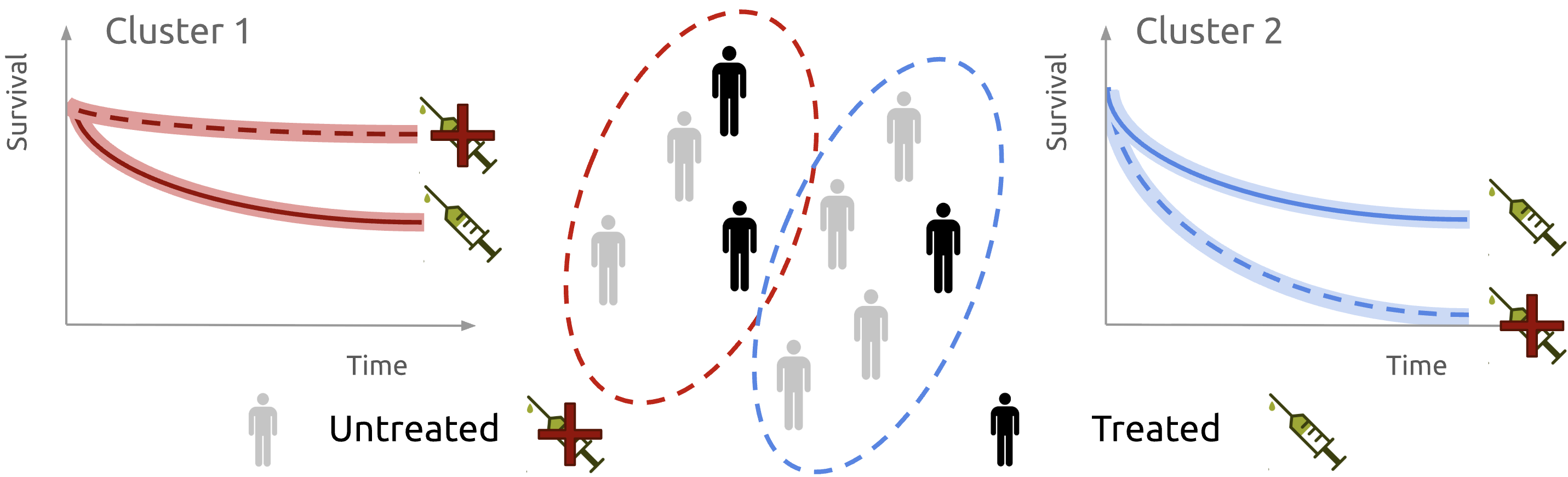}
    \caption{
    \textit{Subgroup treatment effect discovery in time-to-event observational data.}
    Our work aims to identify subgroups of patients with similar treatment responses to guide clinical practice and design clinical trials. 
    Our method simultaneously models the treatment effect and identifies subgroups while addressing treatment non-randomisation and censoring.}
    \label{fig:summary}
    \vspace{-10pt}
\end{figure*}

Understanding heterogeneous therapeutic responses among patient subgroups is central to developing clinical guidelines and new treatments. Identifying such subgroups is valuable to inform the design of future clinical trials and to direct healthcare resources to those most likely to benefit, and away from those at greatest risk of harm~\citep{foster2011subgroup}. A striking example comes from the BARI trial on coronary artery disease, which suggested that patients with diabetes should receive coronary artery bypass grafts rather than percutaneous interventions, whereas patients without diabetes benefited from the opposite strategy~\citep{bypass1996comparison}, an observation that shaped subsequent guidelines. Figure~\ref{fig:summary} illustrates this concept: two groups with opposing treatment responses may require distinct therapeutic recommendations. Our work aims to uncover such subgroups in observational, time-to-event data.

Randomised controlled trials (RCTs) remain the gold standard for studying heterogeneous treatment effects. By randomly assigning patients to control or treated arms, RCTs eliminate potential confounders when assessing a treatment’s impact on outcomes. However, RCTs are often costly, time-consuming, and restricted to specific patient cohorts, which are likely not to reflect the full spectrum of patients seen in clinical practice~\citep{hernan2016using, hernan2010causal, cole2008constructing}. Trial findings may thus not generalise to real-world settings. 

To address these limitations, researchers increasingly leverage large-scale observational datasets which capture diverse, real-world populations. Although observational data offer the potential to identify subgroups that RCTs might miss, they also pose challenges due to non-random treatment assignments and confounding~\citep{benson2000comparison, hernan2010causal, hernan2018estimate}. A robust methodological literature in causal inference has emerged to tackle these challenges. Techniques such as inverse probability weighting~\citep{cole2008constructing}, marginal structural models~\citep{robins2000marginal}, and doubly robust estimators~\citep{bang2005doubly} allow for more accurate causal effect estimation under appropriate assumptions.

Prior works in machine learning (ML) have extended these ideas to estimate treatment effects from observational data~\citep{bica2020estimating, curth2021survite, louizos2017causal}. However, most methods focus on (i)~\textit{average} treatment effects (ATE), which measure the response at the population level, (ii)~\textit{conditional} average treatment effect (CATE), which reflects average response for an individual given its covariates, or (iii)~\textit{individualised} treatment effects (ITE), which aims to estimate the unobservable counterfactual effect. These approaches overlook the utility of identifying \textit{treatment-effect subgroups} ---groups with distinct responses to a given intervention. Identifying such subgroups is crucial in translating treatment effect estimates into actionable insights for clinical decision-making and resource allocation.

We address this gap by introducing a novel framework that uncovers patient subgroups with distinct treatment responses using observational time-to-event data. Extending the rich tradition of outcome-guided modelling~\citep{foster2011subgroup} to a flexible, neural network–based architecture, our proposed approach addresses the limitations of existing methodology for subgroup discovery (see App.~\ref{sec:litreview} for a full review). Crucially, our approach does not require explicit parametrisation of time-to-event distributions or treatment effects; it instead leverages a \textit{mixture of monotonic neural networks} with correction for non-random treatment assignment. In summary, our contributions are:
\begin{itemize}[noitemsep, topsep=0pt]
    \item \textbf{Novel formalisation.} 
    In Sec.~\ref{sec:method}, we formalise the problem of treatment subgroup discovery.
    \item  \textbf{Neural network-based treatment subgroup identification.} We introduce, in Sec.~\ref{sec:method}, a neural network architecture to jointly estimate and identify subgroups of treatment effects in \emph{observational} settings.
    \item \textbf{Extensive application on time-to-event data.} We evaluate its performance on synthetic data in Sec.~\ref{sec:exp}, with extensive sensitivity analysis in App.~\ref{sec:simulation}, and a real-world example described in Sec.~\ref{sec:seer}, benchmarking against several established approaches.
\end{itemize}

\section{Method}
\label{sec:method}

\subsection{Problem setup}

Our goal is to uncover a pre-specified $K \in \mathbb{N}$ number of subgroups\footnote{As a pre-specified number of subgroups may be a limitation in a real-world setting where we do not know the underlying grouping structure, we explore how to select this parameter based on the likelihood of the predicted outcomes in Sec.~\ref{sec:exp}.} with distinct treatment responses, guided by the \emph{observed} times of occurrence of an outcome of interest. 
To this end, our model assigns patients to latent subgroups based on their covariates at the time of treatment decision. For each subgroup, we estimate the probability of not observing the event of interest over time under treatment and under control regimes. These two distributions are known as survival functions, and their difference corresponds to the estimated treatment effect for a given group.

Formally, consider the random variables associated with observed covariates $X$, the assigned \textit{binary} treatment $A$, and the observed event time $T'$. Following the potential outcomes formulation, $T' = A \cdot T'_1 + (1 - A) \cdot T'_0$ under consistency (Asmp.~\ref{assumption:cons}) where $T'_1$ is the potential event time under treatment and $T'_0$, under the control regime. 
\begin{asm}[Consistency]
    \label{assumption:cons}
    A patient's observed event time is the potential event time associated with the observed treatment. Formally, this means $T' = A \cdot T'_1 + (1 - A) \cdot T'_0$ where $T'$ is the observed event time and $(T'_0, T'_1)$ are the potential event times under the control and treatment regimes, respectively.
\end{asm}

Central to our problem is the latent, \textit{unobserved}, subgroup membership $Z$. As formalised in Asmp.~\ref{assumption:mix}, we assume that the variable $Z$ mitigates the dependence between the covariates $X$ and the potential event times. Furthermore, we assume that group membership and treatment assignment are independent given the observed covariates, a plausible assumption because the methodology aims to uncover \textit{unknown} treatment-effect subgroups, as formalised in Asmp.~\ref{assumption:unknown}.
\begin{asm}[Mixture mitigation]
    \label{assumption:mix}
   The event times $(T'_0, T'_1)$ are independent of the covariates given the patient's group membership $Z$. Formally, $(T'_0, T'_1) \indep X \mid Z$.
\end{asm}

\begin{asm}[Unknown groups]
    \label{assumption:unknown}
   The treatment is independent of the group membership given the observed covariates. Formally, $A \indep Z \mid X$.
\end{asm}

\begin{figure}[!h]
    \centering
    \includegraphics[width=0.25\textwidth]{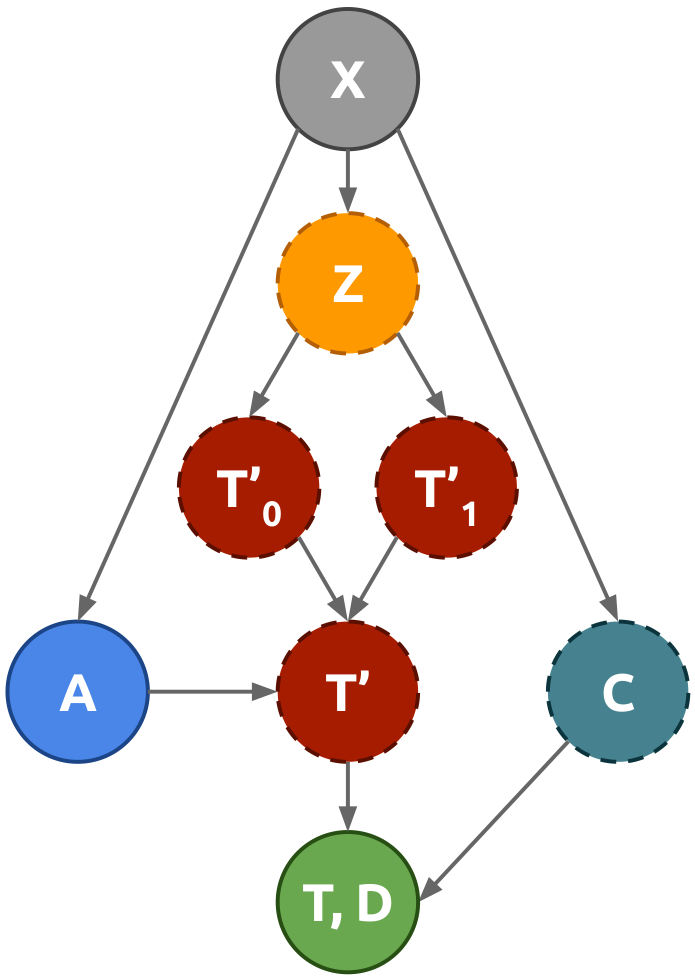}
    \caption{Graphical representation between covariates ($X$), treatment ($A$) and outcomes ($T, D$). Realisations of dashed variables are unobserved, while $X$, $A$, $T$ and $D$ are observed.}
    \label{fig:plate}
    \vspace{-10pt}
\end{figure}

Figure~\ref{fig:plate} summarises all variables and dependencies assumed in the studied problem with a directed acyclic graph. Given the observed $X$, $A$, $T$, and $D$, our aim is to estimate the latent structure $Z$ and, for each group  $k$, the associated \textbf{Subgroup Average Treatment Effect} (SATE):
\begin{align}
\tau_k(t) &= \mathbb{P}(T' \geq t \mid A = 1, Z = k) \notag\\
&- \mathbb{P}(T' \geq t \mid A = 0, Z = k) \label{nsc:equation:subgroup}
\end{align}

This focus differs from the literature interest in estimating CATE, connected to the previous quantity as follows (see App.~\ref{sec:ite} for derivation): 
\begin{align*}
\tau(t, x) 
        &= \sum_k \tau_k(t) \mathbb{P}(Z = k \mid X = x) 
\end{align*}

Estimating the previous quantities requires the accurate modelling of the survival distributions under the two treatment regimes.
The central challenge in this estimation is that the \textit{counterfactual} survival outcome is unobserved: if a patient receives the treatment, we do not observe its outcome under no treatment, and vice versa. 

The absence of treatment randomisation in observational studies, e.g.~clinicians might recommend more aggressive treatment for more severe conditions, results in a covariate shift between the treated and non-treated populations~\citep{curth2021survite} as their covariate distributions differ~\citep{bica2021real}. This prohibits the estimation of the survival functions through the maximisation of the likelihood of the observed outcomes.


Under the common Asmp.~\ref{assumption:ign} and~\ref{assumption:pos}~\citep{hernan2010causal, robins1986new}, 
methods exist to account for the difference between the treated and non-treated populations. One such method adopted in this work is the Inverse Propensity Weighting (IPW), which consists of weighting each observation by the inverse probability of receiving treatment to estimate the overall likelihood.

\begin{asm}[Conditional ignorability]
    \label{assumption:ign}
    The potential event times are independent of the treatment given the observed covariates, i.e. $A \indep (T'_0, T'_1) \mid X$. Equivalently, no unobserved confounders impact both treatment and event time.
\end{asm}

\begin{asm}[Overlap / Positivity] 
    \label{assumption:pos}
    Each patient has a non-zero probability of receiving the treatment, i.e. $\mathbb{P}(A \mid X)\in (0, 1)$ where $(0, 1)$ is the open interval, resulting in a non-deterministic treatment assignment. 
\end{asm}

A final challenge exists in time-to-event data: \textit{censoring}. Patients may not observe the outcome of interest during the study period. 
Formally, instead of observing $T'$, one observes an indicator identifying whether the event of interest was observed $D$, and the associated observed time of event $T$, with $T := \min(C, T')$ and $D := \mathbbm{1}(C > T')$, where $C$ is the random variable of the (right)-censoring time. When a patient is censored, $T = C$ and $D = 0$; otherwise, the event under treatment regime $A$ is observed and $T = T'_A$ and $D = 1$.

As ignoring the censoring process biases the likelihood, we assume, as commonly done in survival analysis, that this process is uninformative, formalised as:
\begin{asm}[Non-informative censoring]
    \label{assumption:cens}
The censoring time $C$ is independent of the time of the event of interest $T'$, given the covariates $X$. Formally, $T' \indep C \mid X$.
\end{asm}

Under the previous assumptions, common in the causal and survival literature, we can maximise the overall likelihood by optimising for the following weighted factual log-likelihood $l_F$:
\begin{align}
    l_F &= \sum_{i, d_i = 1} w_{i} \log \left(- \frac{\partial S(t \mid x_i, a_i)}{\partial t}\bigg|_{t = t_i} \right) \notag\\&+ \sum_{i, d_i = 0} w_{i} \log S(t_i \mid x_i, a_i) \label{eq:loglikelihood}
\end{align}
where $i$ is the patient index, $(x_i, t_i, a_i, d_i)$ are realisations of the associated variables $X, T, A$ and $D$, and $w_i$ is the inverse propensity weighting correction for patient $i$. Under Asmp.~\ref{assumption:mix} and~\ref{assumption:unknown}, 
\begin{align}
S(t &\mid x, a):= \mathbb{P}(T' \geq t \mid x, a) \notag\\&= \sum_{k = 1}^K \mathbb{P}(Z = k \mid x) \mathbb{P}(T' \geq t \mid a, k) \label{nsc:equation:survival}
\end{align}

\subsection{Estimating the quantities of interest}
\label{sec:ntc}
\begin{figure*}[t!]
    \centering
    \includegraphics[width=0.85\textwidth]{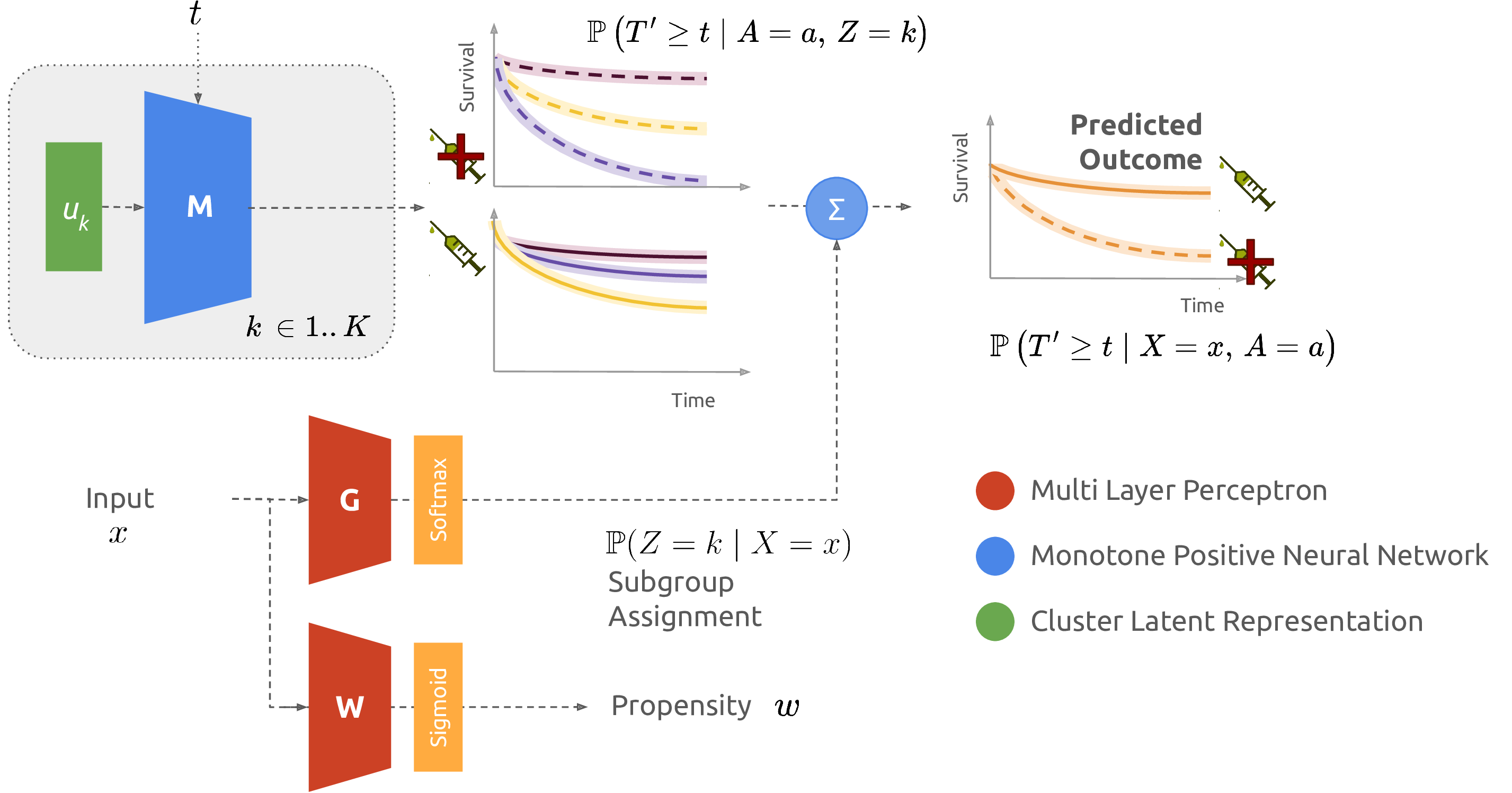}
    \caption{Causal Survival Clustering (CSC) architecture. 
    Latent parameter $u_k$ characterising the subgroup $k$ is inputted into the monotonic network $M$ to estimate the survival under both treatment regimes. $G$ assigns the probability of belonging to each subgroup given the patient's covariate(s) $x$.
    The network $W$ estimates the treatment propensity used to account for the treatment assignment bias.}
    \label{fig:model}
    \vspace{-10pt}
\end{figure*}

The previous section discussed the quantities one must estimate ---here parametrised by neural networks\footnote{When interpretability is a key concern, one can consider a linear model for both subgroup assignment and IPW components, as explored in App.~\ref{sec:linear}.}--- to uncover subgroups of treatment effects: the assignment function $\mathbb{P}(Z\mid X)$, the distributions characterising survival under the two treatment regimes $\mathbb{P}(T' \geq t \mid A, Z = k)$, and the IPW weights $w$. Figure~\ref{fig:model} illustrates the overall architecture and the neural networks used to estimate these quantities.

\paragraph{Subgroup assignment.} A multi-layer perceptron $G$ with a final Softmax layer assigns a patient characterised by covariates $x$ its probability of belonging to each subgroup, characterised through a $K$-dimensional vector of probabilities. 
\begin{align*}
    G(x) := [\mathbb{P}(Z = k \mid x)]_{k = 1}^K
\end{align*}

\paragraph{Survival distributions.}
Each subgroup $k$ is represented by a parameter vector $u_k \in \mathbbm{R}^L$ of dimension $L$, a latent parametrisation learnt through backpropagation. The vector $u_k$ is concatenated with $t$ and used as input to a neural network $M$ with monotonic outcomes as  in~\cite{jeanselme2022neural, jeanselme2023neural} and a final SoftPlus layer to ensure positivity. We train this neural network to model survival through the
following transformation --- ensuring that no probability is assigned to negative times, a limitation raised concerning previous monotonic neural networks~\citep{shchur2019intensity}:
$$\mathbb{P}(T' \geq t \mid A = a, Z = k) := \exp (- t \cdot M(u_k, t)[a])$$

\paragraph{Inverse propensity weighting.} 
As previously discussed, to account for the treatment non-randomisation in observational studies, we weight the factual likelihood using the propensity of receiving treatment estimated through a multi-layer perceptron $W$ with a final sigmoid transformation as:
$$W(x) := \mathbb{P}(A = 1 \mid x)$$

From this estimated probability, we further truncate the propensity score~\citep{austin2015moving} to avoid extreme values for the estimated weights, which would result in unstable estimates of the treatment effect. In this context, the weights $w_i$ are defined as:
\begin{align*}
    \forall i, w_i' :&= a_i W(x_i) + (1 - a_i) (1 - W(x_i))\\ w_i :&= \max(0.05, \min(w_i', 0.95))
    \label{eq:ntc:weights}
\end{align*}

\subsection{Training procedure}
The model $W$ is trained to predict the binary treatment assignment probability by minimising the cross-entropy of receiving treatment. 
Then, all other components are trained by maximising the weighted log-likelihood introduced in Equation~\eqref{eq:loglikelihood}. 
The use of monotonic neural networks results in the efficient and exact computation of the log-likelihood, as a forward path estimates survival and automatic differentiation provides the derivative necessary for computing the log-likelihood~\citep{jeanselme2022neural, jeanselme2023neural, rindt2022survival}.

\section{Experimental analysis}
\label{sec:exp}
As counterfactuals are unknown in observational data, we adopt the common practice of a synthetic dataset in which the underlying survival distributions and group structure are known. 
As a real-world case study, App.~\ref{sec:seer} accompanies these synthetic results with the analysis of heterogeneity in adjuvant radiotherapy responses for patients diagnosed with breast cancer. Our code to produce the synthetic dataset, the model, and all experimental results is available on Github\footnote{\url{https://github.com/Jeanselme/CausalNeuralSurvivalClustering}}.

\subsection{Data generation}
We generate a population of $3,000$ equally divided into $K = 3$ subgroups. 
We draw 10 covariates from normal distributions with different means and survival distributions for each treatment regime following Gompertz distributions~\citep{pollard1992gompertz} parametrised by group membership and individual covariates. Note that this setting breaks Asmp.~\ref{assumption:mix} as survival distributions under treatment or control regimes are different functions of the covariates. This simulation is more likely to capture the complexity of real-world responses, in contrast to traditional evaluations of subgroup analysis, which often assume a linear treatment response.
Further, we implement two treatment assignment scenarios: a randomised assignment in which treatment is independent of patient covariates, similar to RCTs (\textbf{Randomised}), and one in which treatment is a function of the patient covariates as in observational studies (\textbf{Observational}). 
Finally, non-informative censoring times are drawn following a Gompertz distribution. 
Details of the generative process for our synthetic dataset are deferred to App.~\ref{sec:data_generation}. Further, App.~\ref{sec:alternative} explores alternative datasets to demonstrate the robustness of the proposed methodology. 

\subsection{Empirical settings}

\paragraph{Benchmark methods.} We compare the proposed approach (CSC) against the state-of-the-art Cox Mixtures with Heterogeneous Effect (\textit{CMHE}~\cite{nagpal2022counterfactual}\footnote{Implemented in the Auton-Survival library~\citep{autonsurvival}}), which uncovers treatment effect and baseline survival latent groups. 
This method uses an expectation maximisation framework in which each patient is assigned to a group for which a Cox model is then fitted. 
The central differences with our proposed approach are that CMHE (i) clusters patients in treatment effects subgroups and survival subgroups, and (ii) assumes a linear treatment response. This separation between survival ($M$) and treatment response ($L$) enhances the model's flexibility at the expense of interpretability, as the total number of groups grows exponentially ($L \times M$), and subgroups of treatment effect become independent of survival. Further, the assumption of linear treatment response may hinder the discovery of subgroups with more complex responses.
By contrast, our approach identifies treatment subgroups while considering survival without constraining the treatment response. 
We argue that these are key strengths of our method, as a group not responding to treatment with low life expectancy would most benefit from alternative treatments, compared to a group with the same treatment responses but with better survival odds. Considering jointly non-linear treatment effects and survival, therefore, results in identifying more clinically relevant subgroups.

For a fair comparison, we present three alternatives of CMHE: one with fixed $M = 3$ survival subgroups and $L = 1$ treatment subgroups, one with $L = 3$ treatment and $M = 1$ survival subgroups, 
and one with $M = L = 2$, which allows for a total of 4 subgroups. We consider these alternatives to obtain a total number of clusters close to the underlying number of subgroups $K$.
Crucially, CMHE assumes proportional hazards for each subgroup and does not account for the treatment non-randomisation.

As an ablation, we compare our model against its unadjusted alternative (CSC Unadjusted), which uses the unweighted factual likelihood ($w_i = 1$) and therefore is not robust to non-random treatment assignment.

Finally, we compare against two step-wise approaches in which clustering and treatment effect are trained separately. First, we use an unsupervised clustering algorithm on the covariates, followed by a non-parametric estimate of the treatment effect as proposed in~\cite{nagpal2022counterfactual}, referred to as \textit{KMeans + TE} in the following. Specifically, we use a KMeans~\citep{hartigan1979algorithm} to cluster the data, and we compute the difference between Kaplan-Meier~\citep{kaplan1958nonparametric} estimates between control and treated patients stratified by clusters. 
Second, we use a \textit{Virtual Twins} approach to estimate individualised treatment effects and then cluster them. As proposed in the literature, we use a survival tree to estimate response under each treatment regime and then use a KMeans~\citep{macqueen1967some} on the estimated treatment effects, computed as the difference between survival estimates. 
For details on training and hyperparameter optimisation, we refer to App. \ref{sec:gridsearch}.

\paragraph{Evaluation.} In the synthetic experiments, the subgroup structure is known. We measure the adjusted\footnote{Random patient assignment results in an adjusted Rand-Index of 0.} Rand-Index~\citep{rand1971objective}, which quantifies how the estimated subgroup assignment aligns with the generative group structure. 
Additionally, we use the integrated absolute error (IAE) between the treatment effect estimate and the ground truth, which measures how well the methods recover each subgroup's treatment effect
$$\text{IAE}_k(t) = \int_0^t \left|\hat{\tau}_{\hat{k}}(s) - \tau_k(s)\right| ds, $$
where $\hat{\tau}_{\hat{k}}$ is the estimated treatment effect for subgroup $\hat{k} = \argmax_l \mathbb{E}_{x \in k}(Z = l \mid x)$, i.e. the most likely assigned cluster for patients in the underlying $k$ cluster, and $\tau_k$ is the ground truth.

\subsection{Treatment effect recovery}
\label{subsec:treatment}

\begin{figure}[!h]
    \centering
    \includegraphics[width = 0.35\textwidth]{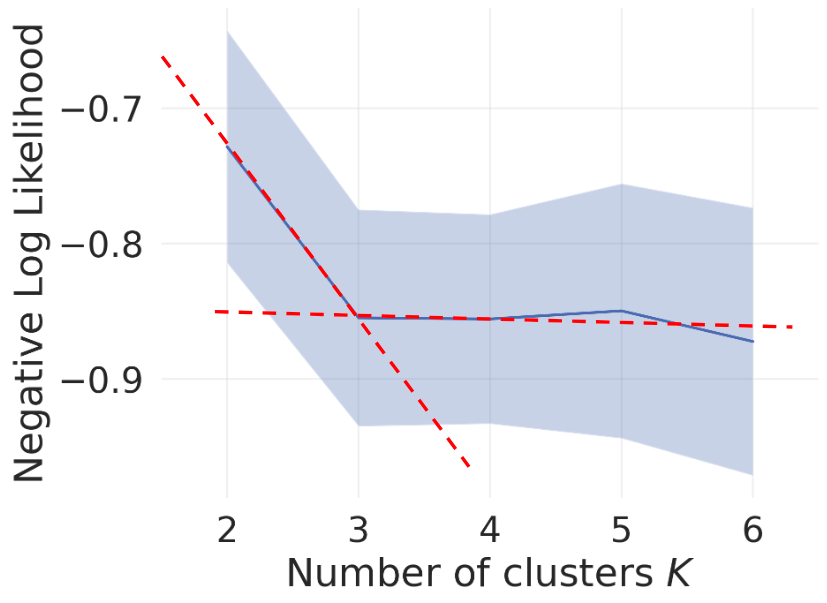}
    \caption{Averaged negative log-likelihood across 5-fold cross-validation given the number of subgroups $K$ under the "Observational" treatment assignment with the shaded area representing 95\% {CI}. \textit{The log-likelihood presents an elbow around the underlying number of subgroups.}}
    \label{ntc:fig:log_k}
    \vspace{-10pt}
\end{figure}

\paragraph{Recovering the underlying number of subgroups.} For all methodologies, we must choose the number of subgroups $K$ \textit{a priori}.
An important question is, therefore, whether we can identify the underlying number of subgroups in a principled way using the proposed CSC.
Figure~\ref{ntc:fig:log_k} presents the average negative log-likelihood obtained by cross-validation for models with different numbers of subgroups $K$. 

The dotted lines represent the elbow heuristic~\citep{thorndike1953belongs}, which identifies a change point in the explained variability, here considering the log-likelihood. 
Using this heuristic, the optimal choice for $K$ is $3$, which aligns with the generative process. App.~\ref{sec:sensitivity} shows that when the number of clusters is misspecified, the estimated treatment effects are unstable; the variance associated with estimated treatment effects is hence an additional tool for informing the choice of $K$.
These data-driven approaches to select the number of subgroups $K$ are a crucial strength of our method, compared to classical two-stage analyses, which separate clustering from treatment effect estimates. In these methods, survival outcomes cannot directly guide the choice of $K$, whereas it is directly reflected in CSC's performance.

\paragraph{Discovering subgroups.} 
Table~\ref{nsc:table:simulation} summarises the performance of the different methodologies under the two studied scenarios. 
Recall that $L$ denotes the number of treatment response subgroups, while in CMHE, the additional parameter $M$ describes the number of survival clusters. 

\begin{table*}[!ht]
    \footnotesize
    \addtolength{\tabcolsep}{-2pt}
    \centerline{
    \begin{tabular}{r|c||c|ccc}
          &  \multirow{2}{*}{Model}  & \multirow{2}{*}{Rand-Index} &\multicolumn{3}{c}{$\text{IAE}_k(t_{max})$}\\
           & & & $k = 1$ & $k = 2$ & $k = 3$\\ \midrule
     \multirow{7}{*}{\rotatebox[origin=c]{90}{Randomised}}
            & \textbf{CSC}     & \textbf{0.807} (0.055) &  \textit{0.037} (0.008) & \textit{0.023} (0.008) &\textit{ 0.015} (0.003) \\
            & \textbf{CSC Unadjusted}   & \textit{0.802} (0.050) &  0.039 (0.011) & \textbf{0.021} (0.009) & \textbf{0.014} (0.003)\\
            & CMHE ($L = 3$)   & 0.392 (0.034) & 0.164 (0.009) & 0.077 (0.004) & 0.070 (0.005) \\
            & CMHE ($M = 3$)   & 0.255 (0.159) & - & 0.084 (0.005) & -\\
            & CMHE ($M = L$)   & 0.345 (0.104) & 0.243 (0.042) & - & - \\
            & KMeans + TE & 0.000 (0.002) & 0.210 (0.011) & 0.091 (0.007) & 0.142 (0.016)\\
            & Virtual Twins & 0.578 (0.081) & \textbf{0.022} (0.006) & 0.032 (0.011) & 0.025 (0.010) \\\midrule
     \multirow{7}{*}{\rotatebox[origin=c]{90}{Observational}} 
            & \textbf{CSC}    &\textbf{0.797} (0.043) & \textbf{0.042} (0.011) & \textit{0.051} (0.026) & \textit{0.022} (0.004)\\
            & \textbf{CSC Unadjusted} &\textit{0.742} (0.073) & \textit{0.047} (0.019) & \textbf{0.030} (0.029) & \textbf{0.020} (0.003) \\
            & CMHE ($L = 3$)  & 0.385 (0.022) &0.169 (0.012) & 0.078 (0.005) & 0.075 (0.008)\\
            & CMHE ($M = 3$)  & 0.190 (0.127) &0.192 (0.010) & - & 0.140 (0.009) \\
            & CMHE ($M = L$)  & 0.454 (0.068) &0.210 (0.013) & 0.095 (0.020) & 0.188 (0.014)\\
            & KMeans + TE & 0.001 (0.004) & 0.192 (0.016) & 0.090 (0.019) & 0.147 (0.016)\\
            & Virtual Twins & 0.438 (0.139) & 0.050 (0.039) & 0.034 (0.033) & 0.051 (0.033)
    \end{tabular}}
    \caption{5-fold cross-validated performance averaged (with standard deviation in parentheses) under the Randomised and Observational treatment simulation. Best performance per column and simulation scenario is marked in \textbf{bold}, second best in \textit{italic}. '-' describes when the methodology diverges. \textit{Our proposed CSC method best recovers the underlying treatment responses in the observational setting.}}
    \label{nsc:table:simulation}
    \vspace{-10pt}
\end{table*}

\textit{CSC outperforms all CMHE alternatives, the current state-of-the-art method.}
CMHE's parametrisation and assumptions explain this difference. Critically, CMHE assumes (i) proportional hazards and (ii) a linear impact of treatment on log hazards. 
Neither of these two assumptions is likely to hold in real-world settings as mimicked by our synthetic data. 
In contrast, our approach does not constrain the treatment effect due to its flexible modelling of the survival function under both treatment regimes. 
Increasing the number of subgroups, as shown in $M = L$, improves the performance of CMHE in terms of clustering quality (Rand-Index) and the recovery of the underlying treatment effect (lower IAE), but still is inferior compared to our proposed methods. These experiments highlight the advantage of the proposed CSC method in uncovering subgroups of treatment responses due to the flexibility in modelling complex survival distributions under both treatment regimes. 

\textit{CSC presents the best performance in identifying the underlying subgroup.}
Our proposed CSC method outperforms all approaches on the Rand-Index. In particular, the two-step approaches do not identify the underlying subgroups when covariates and outcomes subgroups are unaligned. KMeans+TE identifies subgroups that are independent of treatment subgroups, as shown by the Rand-Index and the large IAE. The Virtual~Twins approach presents a better capacity to identify the clusters of interest with better Rand-Index and IAE. However, the two-step approach hurts subgroup identification due to the disconnect between the clustering and input covariates. This disconnect leads to a significantly lower Rand-Index in comparison to CSC despite accurate treatment effect estimates.

\textit{Treatment assignment correction improves subgroup identification in observational settings.}
The Observational simulation demonstrates the importance of correcting the likelihood under treatment non-randomisation. The two CSC alternatives present comparable performance in the randomised setting, as theoretically expected, due to a constant $w_i$ in this context. However, CSC better recovers the different groups as shown by the Rand-Index in the Observational setting.

\section{Case study: Analysis of adjuvant radiotherapy responses}
\label{sec:seer}
To illustrate how practitioners can use the proposed methodology to identify subgroups of treatment response in real-world observational data, we investigate how breast cancer patients respond differently to adjuvant radiotherapy using data from the Surveillance, Epidemiology, and End Results program (\textsc{Seer}). Following \cite{lee2018deephit, danks2022derivative, jeanselme2023neural}, our analysis focuses on women who died from the condition or from cardiovascular diseases. To study the impact of adjuvant radiotherapy on survival outcomes after chemotherapy, we subselect patients with recorded chemotherapy. These criteria led to the selection of 239,855 women with 22 covariates measured at diagnosis, such as diagnosis year, grades, ethnicity, laterality, tumour size, and type (see~\cite{danks2022derivative} for further description).

\begin{table*}[!htb]
\footnotesize
        \addtolength{\tabcolsep}{-1pt}
\centerline{\begin{tabular}{c|ccc||ccc}
    & RMST at 5 years & Population \% & Treated \%& Distant Lymph Nodes & HER2 Positive & ER Positive \\
    \textcolor{blue}{\textbf{Subgroup 0}} & 0.01 (0.00) & 94.6\% &  55.6\% & 1.2 (5.90) & 17.5\% & 46.5\% \\
     \textcolor{red}{\textbf{Subgroup 1}} & 0.84 (0.11) & 5.4\% & 45.1\% & 20.5 (15.56) & 23.4\% & 50.4\% 
\end{tabular}}

\caption{Causal Survival Clustering subgroups' characteristics in the \textsc{Seer} cohort described through percentage / mean (std).}
\label{ntc:tab:result_clusters}
\vspace{-10pt}
\end{table*}

\paragraph{Step 1: Select $K$.}
From this population, our aim is to identify heterogeneous responses to adjuvant treatment. The first challenge is selecting the number of groups to use ($K$). We advise following medical actionability and considering the change in treatment effects and size of the subgroups when increasing this parameter. In the absence of experts' insights, one may rely on the previously described elbow rule heuristic over $l_F$. Using Figure~\ref{fig:seer:loglikelihood} in the Appendix, the negative log-likelihood presents an elbow for $K = 2$. 

\paragraph{Step 2: Identify subgroups.}
Once $K$ is selected, one may use CSC to identify groups of patients who may benefit from adjuvant radiotherapy. This problem is central to patients' management, as no evidence-based guidelines for adjuvant therapy exist~\cite{lazzari2023current}, making this setting more likely to meet the positivity assumption (Asmp.~\ref{assumption:pos}), which is necessary to study causality in observational data.

\begin{figure}[!htb]
    \centering
    \includegraphics[width=0.4\textwidth]{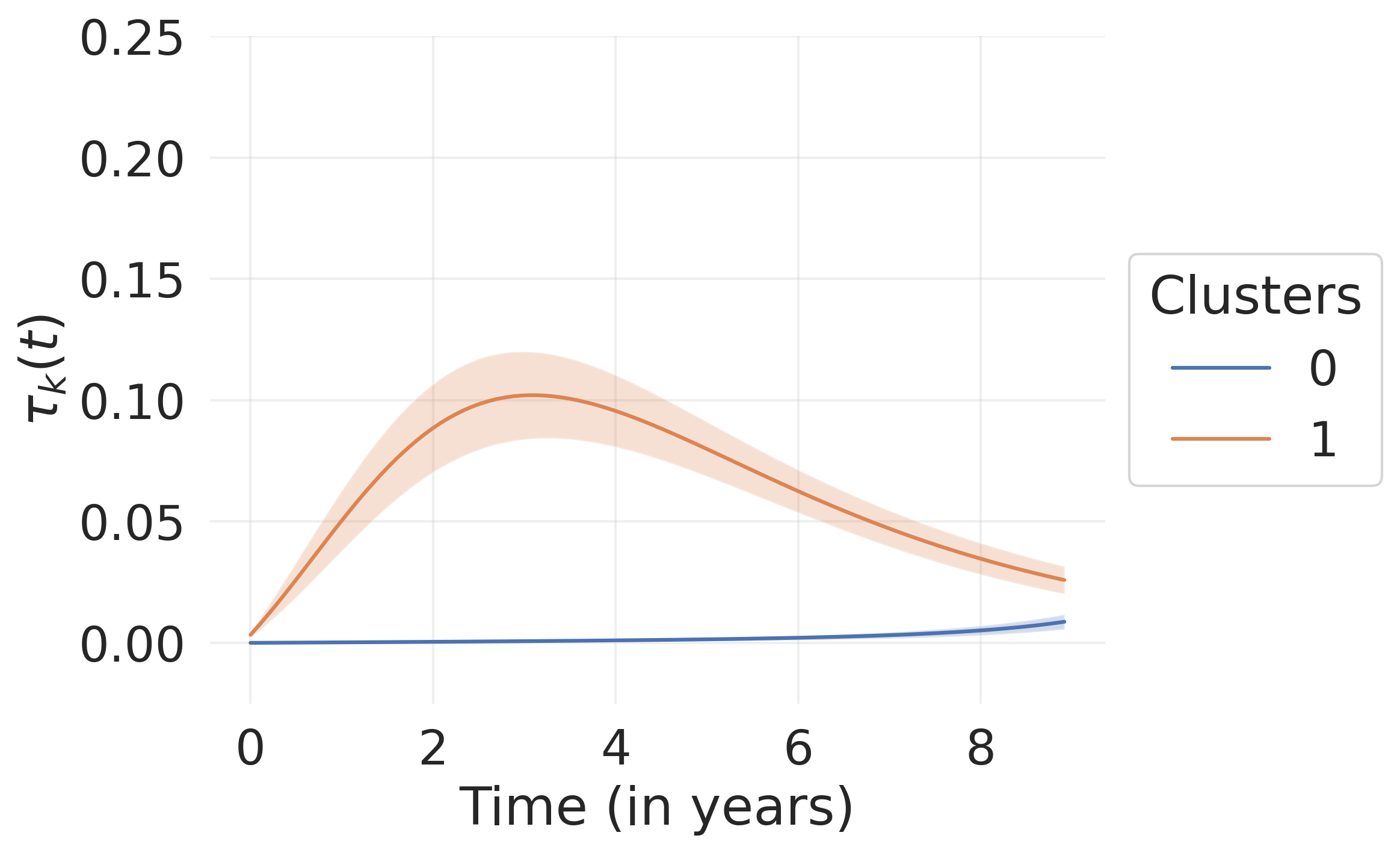}
    \caption{Averaged treatment effect subgroups across 5-fold cross-validation observed in the \textsc{Seer} dataset with the shaded areas representing 95\% CI.}
    \label{ntc:fig:seer:cluster}
    \vspace{-10pt}
\end{figure}

Figure~\ref{ntc:fig:seer:cluster} presents the identified treatment effect subgroups, and Table~\ref{ntc:tab:result_clusters} summarises the mean covariates across the identified subgroups and the life expectancy gain when using adjuvant radiation, measured through the Restricted Mean Survival Time (RMST)~\citep{royston2013restricted}. The proposed methodology identifies two groups: one with limited treatment response and one characterised by larger HER2 prevalence and higher distant lymph node count, with a positive treatment response, gaining more than half a year of life expectancy over the five years following diagnosis. See  App.~\ref{apd:seer}  for the analysis of the treatment subgroups identified using the other baselines.

\paragraph{Step 3: Validate.}
The proposed analysis identifies a group that benefits from adjuvant radiotherapy. However, our methodology remains a hypothesis-generating tool. An important next step would be to develop clinical trials to validate the identified subgroups, particularly due to potential confounding factors such as hormonal therapy (not available in this dataset), the temporal nature of treatment, and the plurality of treatment options. The quality of available covariates limits this analysis and serves only as an example for medical practitioners to identify subgroups of treatment responses from observational data.


\section{Conclusion}
\label{sec:conclusion}
Understanding heterogeneity in treatment effect is essential for guiding clinical practice and informing the development of new treatments. Motivated primarily by the reliance of clinical guidelines on patient subgroups, this work introduces a neural network-based framework for identifying treatment-response subgroups in observational time-to-event data. By leveraging large-scale observational datasets and survival analysis with causal inference methods, our approach addresses a key gap in the literature: it moves beyond average or individualised treatment-effect estimation to reveal subgroups of treatment response.

Our empirical results on both synthetic and real-world datasets demonstrate that the proposed method successfully characterises latent subgroups with divergent treatment effects. While we acknowledge that our model relies on empirically unverifiable assumptions, as is often the case in causal inference, our focus on subgroup detection for hypothesis generation mitigates this concern. Indeed, such subgroup identification serves as a starting point for more targeted RCTs, where further validation can (i)~confirm the heterogeneity in responses, (ii)~explore alternative treatment strategies in less responsive subgroups, and (iii)~shape evidence-based clinical guidelines.

In doing so, this work provides a flexible and practical tool for designing future RCTs. By informing both the selection of patient populations most likely to benefit and avoiding treatments with limited or harmful effects, our framework offers a meaningful hypothesis-generating tool for clinicians, policymakers, and researchers striving to optimise patient care and resource allocation.

\section*{Acknowledgements}
The authors acknowledge the partial support of the UKRI Medical Research Council (programme numbers MC\_UU\_00002/5 and MC\_UU\_00002/2 and theme number MC\_UU\_00040/02 – Precision Medicine).
VJ, CHY and FF acknowledge the Enrichment Scheme of The Alan Turing Institute under the EPSRC Grant EP/N510129/1.
FF acknowledges the receipt of studentship awards from the Health Data Research UK-The Alan Turing Institute Wellcome PhD Programme (Grant Ref: 218529/Z/19/Z).

\bibliography{main}

\clearpage

\appendix
\onecolumn
\section{Related work}
\label{sec:litreview}
While survival subgrouping has been proposed through mixtures of distributions~\citep{nagpal2020deep, nagpal2021deep, jeanselme2022neural} to identify different phenotypes of patients, the ML literature on phenotyping treatment effects with time-to-event outcomes remains sparse. The current focus is on estimating population or individual treatment effects in the static~\citep{shalit2017estimating, johansson2016learning, zhang2020learning} and survival settings~\citep{katzman2018deepsurv, curth2021survite}. While these approaches estimate ATE, CATE, or even ITE, they do not directly identify subgroups that may benefit or be harmed by treatment. Identifying such groups aligns with, and is thus more useful for, drafting medical guidelines that direct treatment to those subgroups most likely to benefit from it. 

Discovering intervenable subgroups is core to medical practice, particularly identifying subgroups of treatment effect, as patients do not respond like the average~\citep{bica2021real, ruberg2010mean, sanchez2022causal} and the average may conceal differential treatment responses. Identification of subgroups has long been used to design RCTs. Indeed, subgroups identified \textit{a priori} can then be tested through trials~\citep{cook2004subgroup, rothwell2005subgroup}. \textit{A posteriori} analyses have gained traction to uncover subgroups of patients from existing RCTs to understand the underlying variability of responses. 

The first set of subgroup analysis methodologies consists of a step-wise approach: (i)~estimate the ITE and (ii)~uncover subgroups using a second model to explain the heterogeneity in ITE. \cite{foster2011subgroup, qi2021explaining} describe the virtual twins approach in which one models the outcome using a decision tree for each treatment group. The difference between these decision trees results in the estimated treatment effects. A final decision tree aims to explain these estimated treatment effects to uncover subgroups. Similar approaches have been explored with different meta-learners~\citep{xu2023treatment}, or Bayesian additive models~\citep{hu2021estimating}, or replacing the final step with a linear predictor to uncover the feature influencing heterogeneity~\citep{chernozhukov2018generic}. However,~\cite{guelman2015uplift} discuss the drawbacks associated with these approaches. Notably, the two-step optimisation may not lead to recovery of the underlying subgroups of treatment effects.

Tree-based approaches were proposed to address the limitations of step-wise approaches by jointly discovering subgroups and modelling the treatment effect. Instead of traditional splits on the observed outcomes, these causal trees aim to discover homogeneous splits regarding covariates and treatment effects. \cite{su2009subgroup} introduce a recursive population splitting based on the average difference in treatment effect between splits. \cite{athey2015machine, athey2016recursive} improve the confidence interval estimation through the honest splitting criterion, which dissociates the splitting from the treatment effect estimation. \cite{wager2018estimation} agglomerate these causal trees into causal forests for improved ITE estimation. Each obtained split in the decision tree delineates two subgroups of treatment effect~\cite{lipkovich2011subgroup, loh2015regression}. Alternatively,~\cite{mcfowland2018efficient} propose pattern detection and~\cite{wang2017causal}, causal rule set learning to uncover these subgroups. However, all these approaches rely on a local optimisation criterion~\citep{lipkovich2014strategies} and greedy split exploration. Recently,~\cite{nagpal2020interpretable} addressed the local optimisation by constraining the treatment response to a linear form in a mixture of Cox models. Our work proposes to address these limitations through a joint optimisation, while avoiding parametric treatment response and proportional hazard assumptions. 

Previous approaches uncover subgroups of treatment effect but consider \textit{RCTs with binary outcomes}, not the observational setting with survival outcomes that our work explores. At the intersection with survival analysis, \cite{zhang2017mining} extend causal trees to survival causal trees, modifying the splitting criterion by measuring the difference in survival estimates between resulting leaves. Similarly, \cite{hu2021estimating} propose Bayesian additive models and \cite{zhu2020targeted} propose a step-wise approach with propensity weighting to study observational data. Closest to our work, \citep{an2023subgroup, jia2021inferring, nagpal2023recovering, perrin2024subgroup} propose to uncover subgroups within RCTs with survival outcomes. 
\cite{perrin2024subgroup} compares different recursive and tree-based appraoches to perform this task in RCTs. \cite{an2023subgroup} introduces an expectation-maximisation approach to jointly fit a logistic model for subgroup attribution and a Cox model for survival estimation. \cite{jia2021inferring} propose a mixture of treatment effects characterised by Weibull distributions trained in an expectation-maximisation framework. Similarly, \cite{nagpal2023recovering} stratify the population into three groups: non-, positive- and negative responders to treatment. An iterative Monte Carlo optimisation is used to uncover these subgroups, characterised by a Cox model with a multiplicative treatment effect. As demonstrated in our work, this step-wise optimisation may be limiting, and the assumption of RCTs renders the model less relevant in observational settings.

\section{Proof}
\label{sec:ite}
This section derives the conditional average treatment effect expression introduced in Sec.~\ref{sec:method}.
\begin{align*}
\tau(t, x) := \mathbb{E}(&\mathbbm{1} (T_1 \geq t) - \mathbbm{1} (T_0 \geq t) \mid X = x)\notag\\
            = \mathbb{E}(&\mathbbm{1} (T_1 \geq t) \mid X = x) 
            - \mathbb{E}(\mathbbm{1} (T_0 \geq t) \mid X = x) \notag\\
            = \mathbb{P}(&T' \geq t \mid A = 1, X = x)
            - \mathbb{P}(T' \geq t \mid A = 0, X = x) \tag{Under Asmp.~\ref{assumption:cons} and~\ref{assumption:ign}}\\
\end{align*}

\section{Synthetic analysis}
\label{sec:simulation}

\subsection{Data generation}
\label{sec:data_generation}
We consider a synthetic population of $N = 3,000$ patients with 10 associated covariates $X \in \mathbb{R}^{10}$ divided into $K = 3$ subgroups. The following data generation does not aim to mimic a particular real-world setting but follows a similar approach to~\cite{nagpal2022counterfactual}. The following describes our generation process:

\paragraph{Covariates.} Each patient's membership $Z$ is drawn from a multinomial with equal probability. Group membership informs the first two covariates through the parametrisation of the bivariate normal distribution with centres $c_k$ equal to $(0, 2.25)$, $(-2.25, -1)$, and $(2.25, -1)$. All other covariates are drawn from standard normal distributions. Formally, this procedure is described as:
\begin{align*}
    Z &\sim \text{Mult}\left(1, \left[\frac{1}{3}, \frac{1}{3}, \frac{1}{3}\right]\right)\\
    X_{[1, 2]} \mid Z = k &\sim \text{MVN}(c_k, I^2)\\
    Y &\sim \text{Mult}\left(1, \left[\frac{1}{4}, \frac{1}{4}, \frac{1}{4}, \frac{1}{4}\right]\right)\\
    X_{[3:10]} \mid Y = k &\sim \text{MVN}(c'_k, I^8)
\end{align*}
with MVN denoting a multivariate normal distribution, $c'_k$ random cluster centres, and $I^n$, the identity covariance matrix of dimension $n$. Note that we introduce $Y$ to present a covariate structure that is independent of the treatment responses.

\begin{figure*}[!ht]
    \subfigure[$K = 2$]{\includegraphics[width=0.32\textwidth]{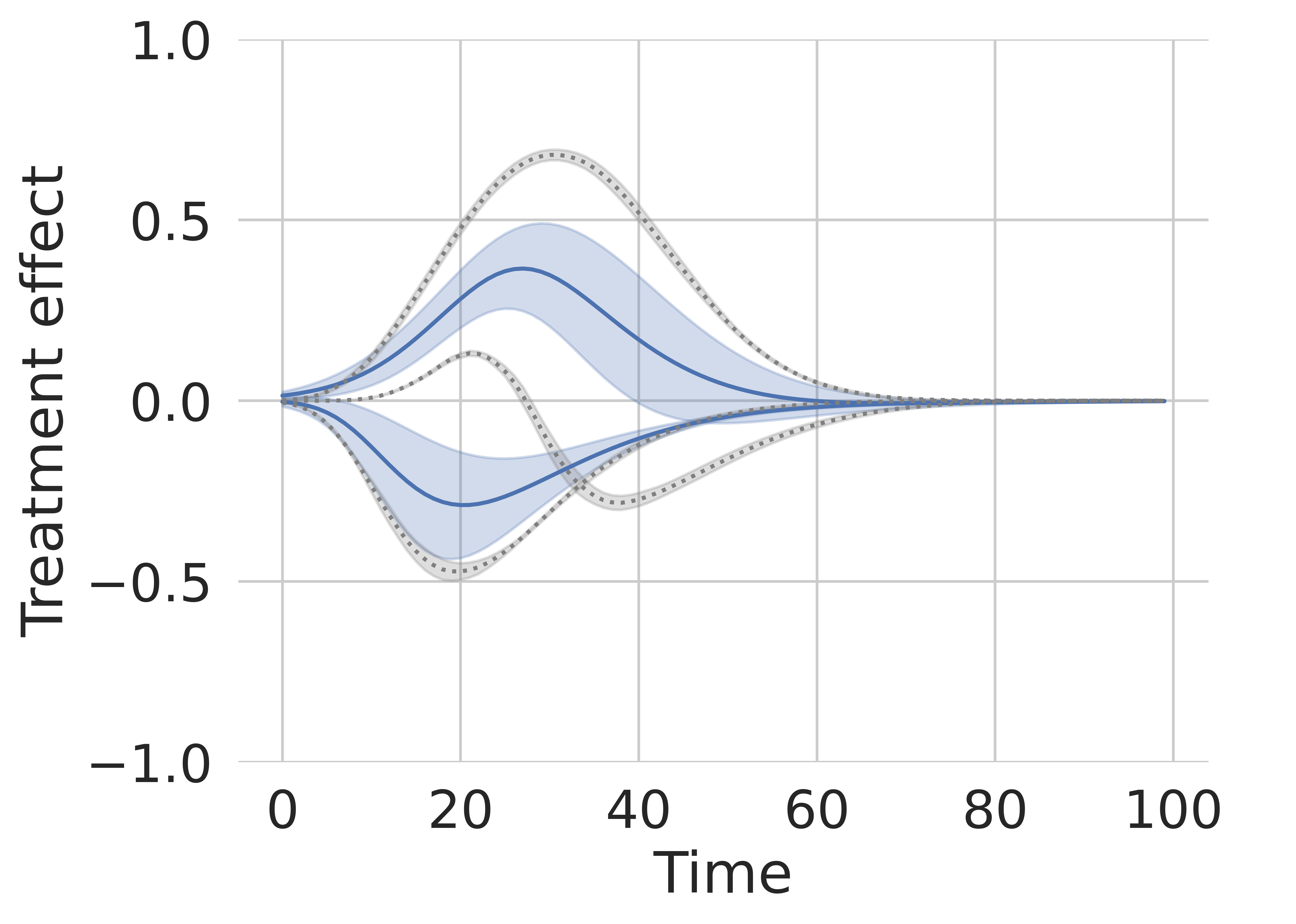}}
    \subfigure[$K = 3$]{\includegraphics[width=0.32\textwidth]{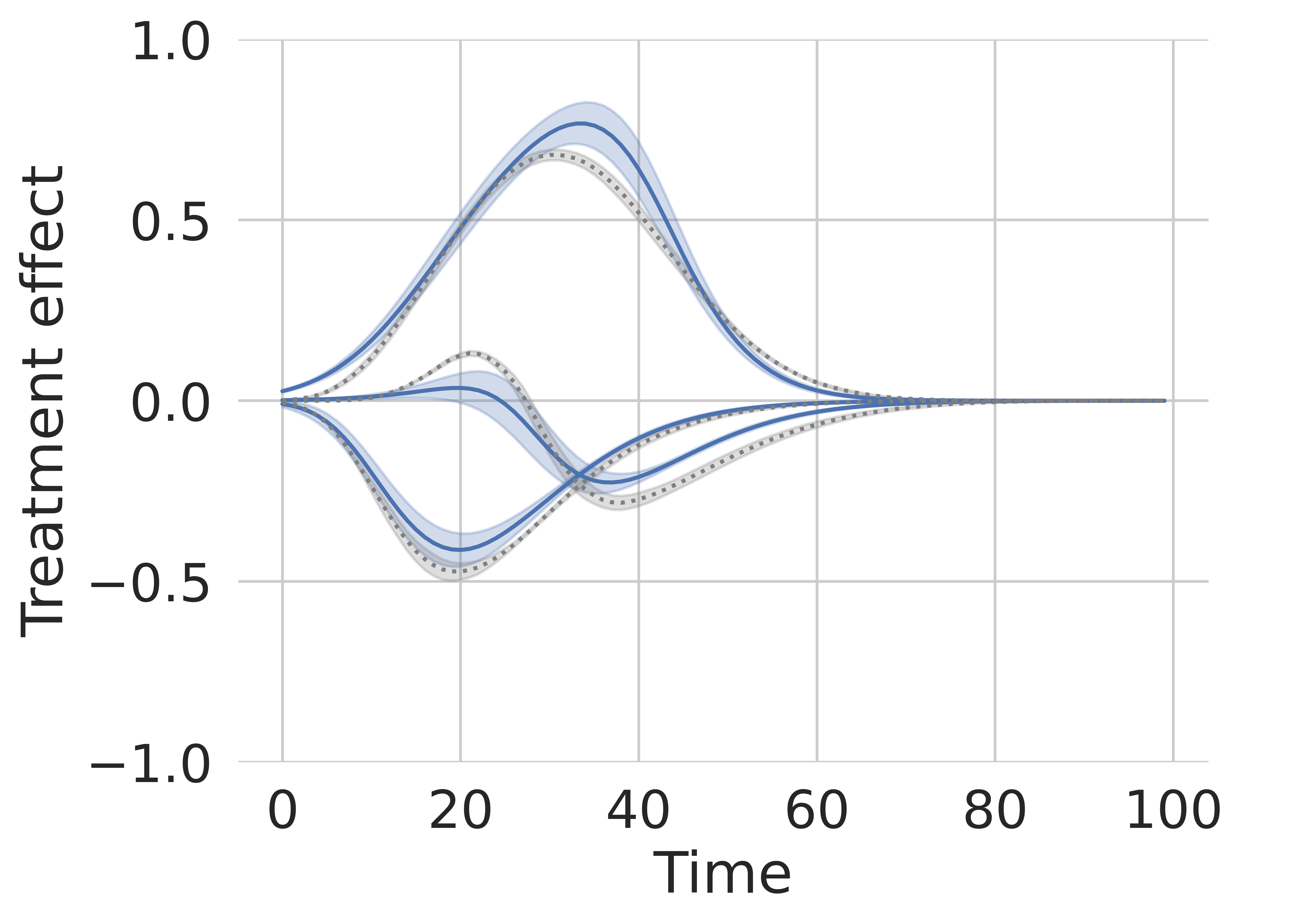}}
    \subfigure[$K = 4$]{\includegraphics[width=0.32\textwidth]{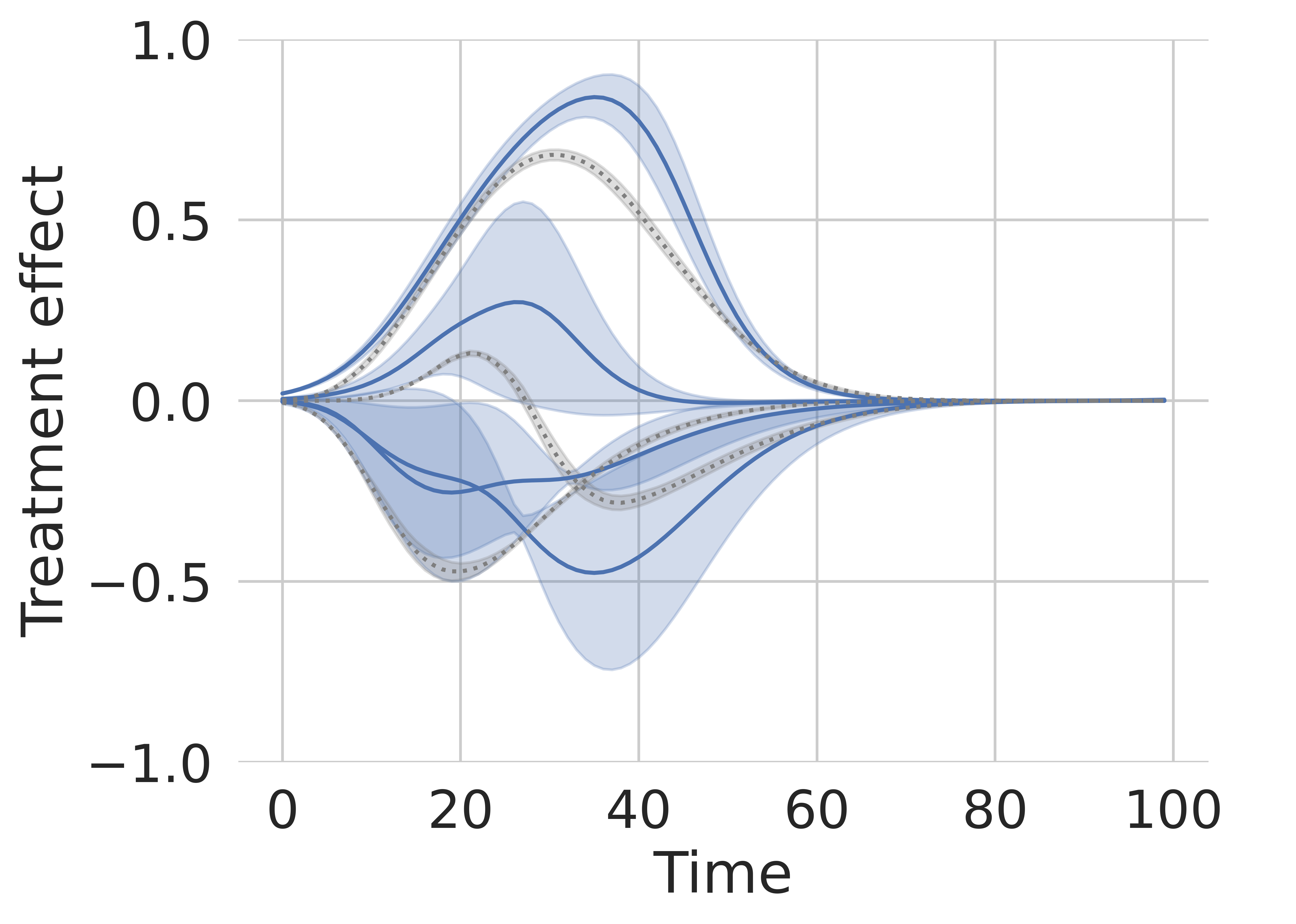}}
  \caption{Sensitivity to the number of clusters used to train CSC. Lines in blue represent the cross-validated average estimated treatment effect. Lines in grey correspond to the ground truth.}
  \label{fig:misspecified}
\end{figure*}

\paragraph{Treatment response.} For each subgroup, event times under treatment and control regimes are drawn from Gompertz distributions, with parameters that are functions of group-specific coefficients ($B^0$ and $\Gamma^0$ for the event time when untreated and $B^1$ and $\Gamma^1$ when treated) and the patient's covariates. 
\begin{align*}
    B^0_z \mid Z = z &\sim \text{MVN}(0, I^{10}) \\
    \Gamma^0_z \mid Z = z &\sim \text{MVN}(0, I^{10}) \\
    T_0 \mid Z, X, B^0_z,& \Gamma^0_z = (z, x, \beta_z^0, \gamma_z^0) \\&\sim \text{Gompertz}\left(w_0(\beta_z^0, x), s_0(\gamma_z^0, x)\right)\\[1.5em]
    B^1_z \mid Z = z &\sim \text{MVN}(0, I^{10})\\
    \Gamma^1_z \mid Z = z &\sim \text{MVN}(0, I^{10})\\
    T_1 \mid Z, X, B^1_z,& \Gamma^1_z = (z, x, \beta_z^1, \gamma_z^1) \\&\sim \text{Gompertz}\left(w_1(\beta_z^1, x), s_1(\gamma_z^1, x)\right)
\end{align*}
with $w_0$, $w_1$ two functions parametrising the Gompertz distributions' shape as $w_0(\beta, x) := |\beta[0]| + (x[5:10] \cdot \beta[5:10])^2$, $w_1(\beta, x) := |\beta[0]| + (x[1:5] \cdot \beta[1:5])^2$, and the shift parameter parametrised as $s_0(\gamma, x) := |\gamma[0]| + |(x[1:5] \cdot \gamma[1:5])| $ and $s_1(\gamma, x) := |\gamma[0]| + |(x[5:10] \cdot \gamma[5:10])|$ where $v[i]$ described the $i^{th}$ element of the vector $v$. These functions aim to introduce non-linear responses with discrepancies between control and treatment regimes. Note that we allow covariates to influence the survival distribution as a patient's covariates influence Gompertz's shapes and scales. 

\paragraph{Treatment assignment.} The non-randomisation of treatment is central to the problem of identifying treatment subgroups in real-world studies. Assuming a treatment assignment probability of 50\%, we assign each patient to a given treatment. We propose two treatment assignment strategies reflecting an RCT and an observational setting, denoted as \textit{Randomised} and \textit{Observational}. \textit{Randomised} consists of a Bernoulli draw using the realisation of $P$. \textit{Observational} reflects an assignment dependent upon the observed covariates.
\begin{align*}
    A_{rand} &\sim \text{Bernoulli}(0.5)\\[1.5em]
    A_{obs} \mid X = x &\sim \text{Bernoulli}(F_{\Phi(X)}(\Phi(x)) \times 0.5)
\end{align*}
with $F_{\Phi(X)}(\Phi(x))$ the cumulative distribution function that returns the probability that a realisation of a $\Phi(X)$ will take a smaller value than $\Phi(x)$. In our experiment, we chose $\Phi(x) =\sum x^2$ for a non-linear treatment assignment.

\paragraph{Censoring.} Finally, our work focuses on right-censored data. To generate censoring independent of the treatment and event, we draw censoring from another Gompertz distribution as follows:
\begin{align*}
    B^C &\sim \text{MVN}(0, I^{5})\\
    C \mid X, B^C = (x, \beta) &\sim \text{Gompertz}\left(w_c(\beta, x), 0\right)\\
    T' &= A \cdot T_1 + (1 - A) \cdot T_0\\
    T &= \min(C, T')\\
    D &= \mathbbm{1}(C > T')
\end{align*}
with $w_c := (x[5:10] \cdot \beta) ^2$, the scale of the censoring Gompertz distribution.

Our goal is to model the treatment effect and identify the underlying subgroup structure $Z$ from the observed $X, T, D, A$ with $T \mid A = a, T_0, T_1, C = t_0, t_1, c := \min(c, (1 - a) \times t_0 + a \times t_1)$ and $D = \mathbbm{1}_{C > T}$.

\subsection{Training and hyperparameter optimisation}
\label{sec:gridsearch}

\paragraph{Training.} 
We perform a 5-fold cross-validation for both Randomised and Observational simulations.
For each cross-validation split, the development set is divided into three parts: 80\% for training, 10\% for early stopping, and 10\% for hyper-parameter search. 
All models were optimised for 1000 epochs using an Adam optimiser~\citep{kingma2014adam} with early stopping.

\paragraph{Hyperparameter optimisation. }
We adopted a 500-iteration random grid-search over the following hyperparameters: network depth between 0 and 3 inner layers with 50 nodes, latent subgroup representation in $[10, 25, 50]$ and for training, a learning rate of $0.001$ or $0.0001$ with batch size of $100$ or $250$. 

The same set of parameters, when appropriate, was used for CMHE. Tree-based approach to estimate treatment effect had a depth limited to: 3, 5 or unconstrained, and the minimum sample split was chosen between 6, 12 or 24.

\subsection{Sensitivity to choice of K}
\label{sec:sensitivity}
In Sec.~\ref{sec:exp}, we describe how the proposed methodology presents an elbow in the likelihood as a function of the number of clusters around the underlying number of clusters $K = 3$. While this outcome-guided selection of the number of clusters is a core strength of the proposed methodology in comparison to existing strategies, we explore how the identified treatment effects change under a misspecified model that estimates 2 and 4 clusters. This analysis examines the risk associated with misspecifying the number of clusters.

\begin{table*}[!ht]
    \small
    \addtolength{\tabcolsep}{-2pt}
    \centerline{
    \begin{tabular}{r|c||c|ccc}
          &  \multirow{2}{*}{Model}  & \multirow{2}{*}{Rand-Index} &  \multicolumn{3}{c}{$\text{IAE}_k(t_{max})$}\\
           & & & $k = 1$ & $k = 2$ & $k = 3$\\ \midrule
     \multirow{3}{*}{\rotatebox[origin=c]{90}{Obs.}} 
            & \textbf{CSC}    &{0.797} (0.043) & \textbf{0.042} (0.011) & {0.051} (0.026) & {0.022} (0.004)\\
            & \textbf{CSC Unadjusted} &{0.742} (0.073) & 0.047 (0.019) & \textbf{0.030} (0.029) & \textbf{0.020} (0.003) \\
            & \textbf{CSC Linear} & \textbf{0.819} (0.041)	 & 0.047 (0.015) & {0.031} (0.022) & {0.022} (0.003)\\
    \end{tabular}}
    \caption{Cross-validated performance (with standard deviation in parentheses).}
    \label{table:logregression}
\end{table*}

\begin{table*}[!ht]
    \small
    \addtolength{\tabcolsep}{-2pt}
    \centerline{
    \begin{tabular}{r|c||c|ccc}
          &  \multirow{2}{*}{Model}  & \multirow{2}{*}{Rand-Index} &  \multicolumn{3}{c}{$\text{IAE}_k(t_{max})$}\\
           & & & $k = 1$ & $k = 2$ & $k = 3$\\ \midrule
     \multirow{7}{*}{\rotatebox[origin=c]{90}{Randomised}}
            & \textbf{CSC}    & \textbf{0.643} (0.067) & \textit{0.060} (0.016) & \textit{0.021} (0.007) & \textit{0.070} (0.013)\\
            & \textbf{CSC Unadjusted} & \textbf{0.643} (0.071) & 0.064 (0.013) & \textbf{0.018} (0.008) & \textbf{0.060} (0.010)\\
            & CMHE ($L = 3$)  & \textit{0.589} (0.019) & 0.179 (0.011) & 0.096 (0.008) & 0.086 (0.004) \\
            & CMHE ($M = 3$)   & 0.072 (0.051) & 0.208 (0.004) & 0.211 (0.027) & 0.122 (0.011)   \\
            & CMHE ($M = L$)   & 0.471 (0.022) & 0.241 (0.011) & 0.264 (0.016) & 0.173 (0.010) \\
            & KMeans + TE & 0.336 (0.010) & 0.078 (0.008) & 0.029 (0.007) & 0.237 (0.017)\\
            & Virtual Twins & 0.541 (0.163) & \textbf{0.033} (0.025) & 0.052 (0.018) & 0.087 (0.051)\\\midrule
     \multirow{7}{*}{\rotatebox[origin=c]{90}{Observational}} 
            & \textbf{CSC}    & \textbf{0.788} (0.113) & \textbf{0.040} (0.017) & \textit{0.041} (0.017) & \textit{0.086} (0.043)\\
            & \textbf{CSC Unadjusted} & \textit{0.626} (0.170) & 0.080 (0.015) & \textbf{0.031} (0.009) & 0.095 (0.049)\\
            & CMHE ($L = 3$)  & {0.611} (0.022) & 0.179 (0.014) & 0.103 (0.005) & {0.089} (0.005) \\
            & CMHE ($M = 3$)   & 0.084 (0.034) & 0.211 (0.007) & 0.209 (0.011) & 0.127 (0.009)\\
            & CMHE ($M = L$)   & 0.462 (0.062) & 0.252 (0.012) & 0.286 (0.008) & 0.096 (0.010)\\
            & KMeans + TE & 0.341 (0.020) & \textit{0.045} (0.009) & 0.043 (0.009) & 0.275 (0.017)\\
            & Virtual Twins & 0.534 (0.166) & \textbf{0.040} (0.015) & 0.051 (0.024) & \textbf{0.084} (0.035)
    \end{tabular}}
    \caption{Cross-validated performance (with standard deviation in parentheses) when clusters are of different sizes.}
    \label{table:size}
\end{table*}

Figure~\ref{fig:misspecified} illustrates the cross-validated subgroup treatment effects when the model is trained with $K = 2, 3$, and $4$ clusters, despite the existence of 3 clusters. When using a number of clusters different from that indicated by the elbow heuristic, the identified clusters are less stable, as they do not capture the true underlying treatment responses, as shown by the increased standard deviation. 
Uncertainty increases with misspecified models, providing an additional tool to select the appropriate number of clusters: the stability of the estimated clusters over cross-validation indicates a better alignment with the underlying distribution.

\subsection{Linear modelling}
\label{sec:linear}
While the flexibility of neural networks allows flexibility of the survival distribution, one could consider more interpretable assignment $G$ and treatment $W$ components than the neural network approach proposed in the main text. This section explores the recovery of treatment effects and subgroups when using linear regressions, i.e. no hidden layers in a neural network, instead of the previous neural networks.

Table~\ref{table:logregression} summarises the results presented in the main paper with the additional linear alternative in which both the treatment and assignment networks are replaced by a linear model, referred as CSC Linear. These results demonstrate that enforcing a linear relation between covariates and the different quantities of interest can reduce the risk of overfitting in small datasets, as shown by the improved Rand-Index. Note that under a larger dataset or a more complex relation between covariates, group membership, and treatment, one should consider more flexible modelling, such as the proposed neural approach.

\subsection{Alternative data generations}
\label{sec:alternative}
 In this section, we explore alternative data-generative processes to demonstrate the robustness of the proposed strategy under different settings.

\subsubsection{Unequal cluster size}
\label{sec:Unequal cluster size}

Sec.~\ref {sec:data_generation} presents an equally distributed population over the different clusters. In medical settings, the underlying subgroups may differ in size. This section explores an alternative scenario in which the population is distributed over the three clusters as follows: 62.5\%, 25\%, and 12.5\%.

Table~\ref{table:size} summarises the performance in this simulation, echoing the main text's conclusions. The proposed method best recovers the different subgroups and presents one of the best estimates of subgroups' treatment effects.

\subsubsection{Number of points}
Sec.~\ref{sec:data_generation} describes a data generation with $3,000$ patients. This section presents results for $300$ and $30,000$ patients. A smaller number of patients may impact the neural network capacity to identify underlying subgroups of treatment effect. A larger number makes non-randomisation less of a concern.

\begin{table*}[!ht]
    \small
    \addtolength{\tabcolsep}{-2pt}
    \centerline{
    \begin{tabular}{r|c||c|ccc}
          &  \multirow{2}{*}{Model}  & \multirow{2}{*}{Rand-Index} &  \multicolumn{3}{c}{$\text{IAE}_k(t_{max})$}\\
           & & & $k = 1$ & $k = 2$ & $k = 3$\\ \midrule
     \multirow{7}{*}{\rotatebox[origin=c]{90}{300}}
            & \textbf{CSC} & \textit{0.310} (0.187) & {0.236} (0.103) & \textit{0.110} (0.033) & {0.149} (0.106)\\
            & \textbf{CSC Unadjusted} & \textbf{0.443} (0.285) & \textit{0.176} (0.098) & \textbf{0.087} (0.046) & \textit{0.132} (0.112) \\
            & CMHE ($L = 3$)  & 0.113 (0.100) & 0.296 (0.040) & \textit{0.137} (0.042) & 0.188 (0.080) \\
            & CMHE ($M = 3$)   & 0.131 (0.067) & - & - & 0.254 (0.022) \\
            & CMHE ($M = L$)   & 0.079 (0.119) & 0.359 (0.068) & 0.182 (0.089) & 0.240 (0.058) \\
            & KMeans + TE & 0.050 (0.110) & 0.251 (0.072) & 0.154 (0.065) & 0.267 (0.049)\\
            & Virtual Twins & 0.296 (0.115) &  \textbf{0.159} (0.072) & 0.172 (0.067) & \textbf{0.101} (0.043)\\\midrule
     \multirow{7}{*}{\rotatebox[origin=c]{90}{30,000}} 
            & \textbf{CSC}    & \textbf{0.764} (0.015) & \textit{0.027} (0.008) & {0.011} (0.001) & 0.047 (0.008)\\
            & \textbf{CSC Unadjusted}  & \textit{0.729} (0.022) & 0.033 (0.008) & \textit{0.010} (0.003) & \textit{0.039} (0.009) \\
            & CMHE ($L = 3$)   & 0.567 (0.159) & 0.094 (0.011) & 0.059 (0.018) & 0.142 (0.034) \\
            & CMHE ($M = 3$)    & 0.622 (0.040) & 0.078 (0.001) & 0.125 (0.003) & 0.226 (0.003)\\
            & CMHE ($M = L$)    & 0.638 (0.008) & 0.059 (0.001) & 0.220 (0.007) & 0.140 (0.007)\\
            & KMeans + TE  & 0.000 (0.000) & 0.087 (0.004) & 0.147 (0.008) & 0.208 (0.004)\\
            & Virtual Twins  & 0.650 (0.030) & \textbf{0.011} (0.003) & \textbf{0.009} (0.000) & \textbf{0.010} (0.004)
    \end{tabular}}
    \caption{Cross-validated performance (with standard deviation in parentheses) with varying $N$ under an observational treatment setting.}
    \label{table:points}
\end{table*}

Table~\ref{table:points} presents the performance with these different population sizes under an observational treatment assignment. A first observation is that all methodologies present better performance with a larger number of points. Further, baselines that do not account for the assignment mechanisms present lower performance with $N = 300$ as non-randomisation has an increased impact on treatment effect estimates with a smaller population. This difference decreases when $N = 30,000$ with the virtual twins approach presenting the best recovery of the treatment effects. However, throughout the different settings, the proposed CSC presents the best Rand-Index, indicating a good recovery of the underlying structure.

\subsubsection{Impact of treatment rate}
Sec.~\ref{sec:data_generation} assumes 50\% of the population receives treatment. This section explores when 25\% and 75\% of the population receive treatment under the non-randomised treatment setting.

Table~\ref{table:treat} presents the performance under these different treatment rates in an observational setting. These results highlight CSC's capacity to identify the underlying subgroups with the highest Rand-Index and one of the best cluster treatment effect recovery in these settings.

\begin{table*}[!ht]
    \small
    \addtolength{\tabcolsep}{-2pt}
    \centerline{
    \begin{tabular}{r|c||c|ccc}
          &  \multirow{2}{*}{Model}  & \multirow{2}{*}{Rand-Index} &  \multicolumn{3}{c}{$\text{IAE}_k(t_{max})$}\\
           & & & $k = 1$ & $k = 2$ & $k = 3$\\ \midrule
     \multirow{7}{*}{\rotatebox[origin=c]{90}{25\%}} 
            & \textbf{CSC}    & \textit{0.830} (0.047)	 & 0.051 (0.008) & \textit{0.023} (0.013) & \textit{0.020} (0.005)\\
            & \textbf{CSC Unadjusted} & \textbf{0.857} (0.033) & \textit{0.045} (0.006) & 0.030 (0.015) & \textbf{0.015} (0.003)\\
            & CMHE ($L = 3$)  & 0.376 (0.028) & 0.168 (0.007) & 0.077 (0.002) & 0.073 (0.004) \\
            & CMHE ($M = 3$)   & 0.134 (0.090) & 0.208 (0.028) & 0.090 (0.015) & 0.134 (0.030) \\
            & CMHE ($M = L$)   & 0.410 (0.038) & 0.223 (0.011)  & -  & -\\
            & KMeans + TE & 0.001 (0.006) & 0.187 (0.005) & 0.088 (0.014) & 0.149 (0.011)\\
            & Virtual Twins & 0.527 (0.060) & \textbf{0.034} (0.012) & \textit{0.019} (0.005) & 0.034 (0.012)\\\midrule
     \multirow{7}{*}{\rotatebox[origin=c]{90}{75\%}} 
            & \textbf{CSC}    & \textit{0.718} (0.198) & 0.061 (0.050) & 0.057 (0.048) & \textbf{0.023} (0.005)\\
            & \textbf{CSC Unadjusted} & \textbf{0.804} (0.045) & \textit{0.037} (0.006) & \textit{0.023} (0.008) & \textit{0.025} (0.007)\\
            & CMHE ($L = 3$)  & 0.385 (0.039) &0.168 (0.007) & 0.077 (0.002) & 0.073 (0.004) \\
            & CMHE ($M = 3$)   & 0.318 (0.140) &  0.208 (0.028) & 0.090 (0.015) & 0.134 (0.030) \\
            & CMHE ($M = L$)   & 0.410 (0.038) & 0.223 (0.011) & - & - \\
            & KMeans + TE & 0.001 (0.006) & 0.187 (0.005) & 0.088 (0.014) & 0.149 (0.011)\\
            & Virtual Twins & 0.527 (0.060)	 & \textbf{0.034} (0.012) & \textbf{0.019} (0.005) & 0.034 (0.012)
    \end{tabular}}
    \caption{Cross-validated performance (with standard deviation in parentheses) with varying treatment rates under observational settings.}
    \label{table:treat}
\end{table*}

\subsubsection{Aligned covariates and treatment effect structure}
\label{sec:Aligned clusters and outcomes}

Sec.~\ref {sec:data_generation} presents a data generation with the last 8 covariates having a clustered structure independent of the treatment response. This section explores a setting where all clusters in the covariates are associated with different treatment responses of interest. Specifically, we sample the last dimensions from standard normal distributions instead of various clusters, i.e.,
$$X_{[3-10]} \sim \text{MVN}(0, I^{8})$$

\begin{table*}[!ht]
    \small
    \addtolength{\tabcolsep}{-2pt}
    \centerline{
    \begin{tabular}{r|c||c|ccc}
          &  \multirow{2}{*}{Model}  & \multirow{2}{*}{Rand-Index} &   \multicolumn{3}{c}{$\text{IAE}_k(t_{max})$}\\
           & & & $k = 1$ & $k = 2$ & $k = 3$\\ \midrule
     \multirow{7}{*}{\rotatebox[origin=c]{90}{Randomised}}
            & \textbf{CSC}     & \textit{0.559} (0.114) & \textit{0.024} (0.009) & \textit{0.030} (0.017) & \textit{0.030} (0.006)\\
            & \textbf{CSC Unadjusted}   & 0.481 (0.069) & 0.026 (0.005) & 0.041 (0.026) & 0.032 (0.009)\\
            & CMHE ($L = 3$)   & 0.425 (0.020) & 0.059 (0.004) & {0.034} (0.025) & 0.134 (0.012)\\
            & CMHE ($M = 3$)   & 0.165 (0.107) &0.062 (0.004) & 0.126 (0.175) & 0.233 (0.179)\\
            & CMHE ($M = L$)   & 0.254 (0.138) & 0.074 (0.027) & 0.052 (0.021) & 0.149 (0.018) \\
            & KMeans + TE & \textbf{0.888} (0.020) &\textbf{0.015} (0.007) & \textbf{0.021} (0.007) & \textbf{0.020} (0.003)\\
            & Virtual Twins & 0.200 (0.090) & 0.035 (0.011) & 0.065 (0.025) & 0.057 (0.030)\\\midrule
     \multirow{7}{*}{\rotatebox[origin=c]{90}{Observational}} 
            & \textbf{CSC}     & \textit{0.584} (0.151) & \textit{0.015} (0.007) & \textbf{0.028} (0.024) & \textit{0.035} (0.003) \\
            & \textbf{CSC Unadjusted}   & 0.556 (0.249)  & 0.023 (0.008) & 0.051 (0.025) & 0.040 (0.006)\\ 
            & CMHE ($L = 3$)   & 0.392 (0.062) & 0.059 (0.008) & \textit{0.034} (0.019) & 0.131 (0.015) \\
            & CMHE ($M = 3$)   & 0.133 (0.027) & 0.059 (0.003) & 0.051 (0.007) & 0.153 (0.004) \\
            & CMHE ($M = L$)   & 0.293 (0.125) & 0.063 (0.012) & 0.051 (0.017) & 0.140 (0.011) \\
            & KMeans + TE& \textbf{0.895} (0.021) & \textbf{0.013} (0.005) & 0.042 (0.010) & \textbf{0.020} (0.004) \\
            & Virtual Twins & 0.226 (0.069) & 0.045 (0.024) & 0.042 (0.031) & 0.048 (0.011)
    \end{tabular}}
    \caption{Cross-validated performance (with standard deviation in parentheses) when all covariate clusters present different treatment responses. This setting aligns with the assumptions made by the step-wise approaches.}
    \label{table:same}
\end{table*}

Table~\ref{table:same} summarises the performance in this setting, evidencing an improvement of the KMeans+TE approach as this model makes this assumption. When covariates' clusters are aligned with the outcome of interest, this method recovers the clustering structure well, as shown by the best Rand-Index.  

Even in this unrealistic scenario, our proposed method remains the second best in recovering the underlying clustering structure and associated treatment responses. This observation reinforces the main findings of our work, demonstrating the method's potential to identify subgroups of treatment effects across different and even adversarial settings.

\section{Additional results: Heterogeneity of adjuvant radiotherapy responses in the \textsc{Seer} dataset}
\label{apd:seer}

This section presents additional results on the heterogeneity of adjuvant radiotherapy responses in the \textsc{Seer} dataset.

\subsection{Selection of $K$}
\begin{figure}[!htb]
    \centering
    \includegraphics[height = 125 px]{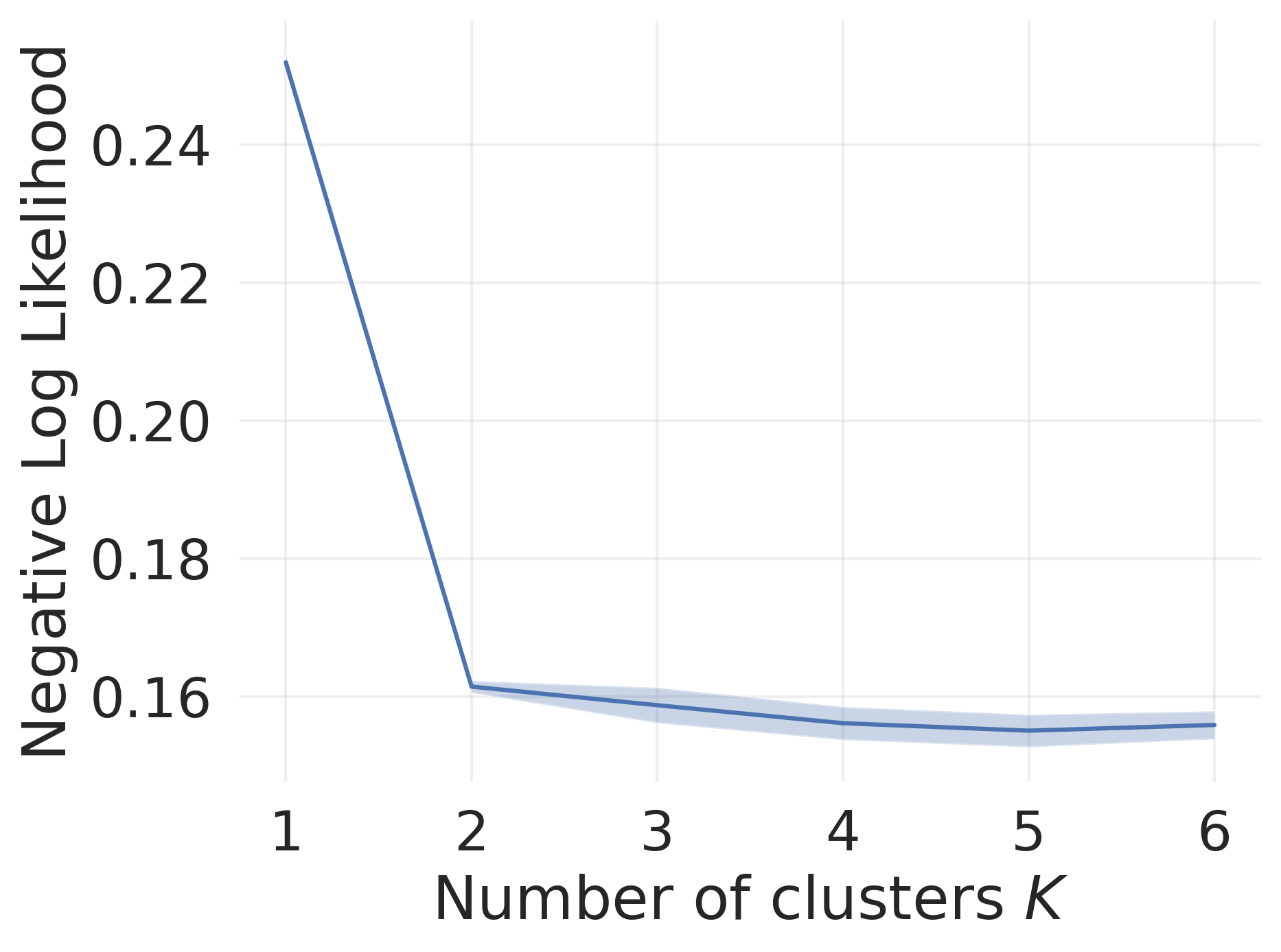}
    \caption{Cross-validated negative log-likelihood as a function of the number of groups ($K$).}
    \label{fig:seer:loglikelihood}
\end{figure}

Using Figure~\ref{fig:seer:loglikelihood}, one can use the elbow rule on the negative log-likelihood. Following this heuristic, our analysis uses $K = 2$ subgroups.

\subsection{Factors impacting subgroup membership}
\begin{figure*}[!htb]
    \centering
    \includegraphics[width = 0.5\textwidth]{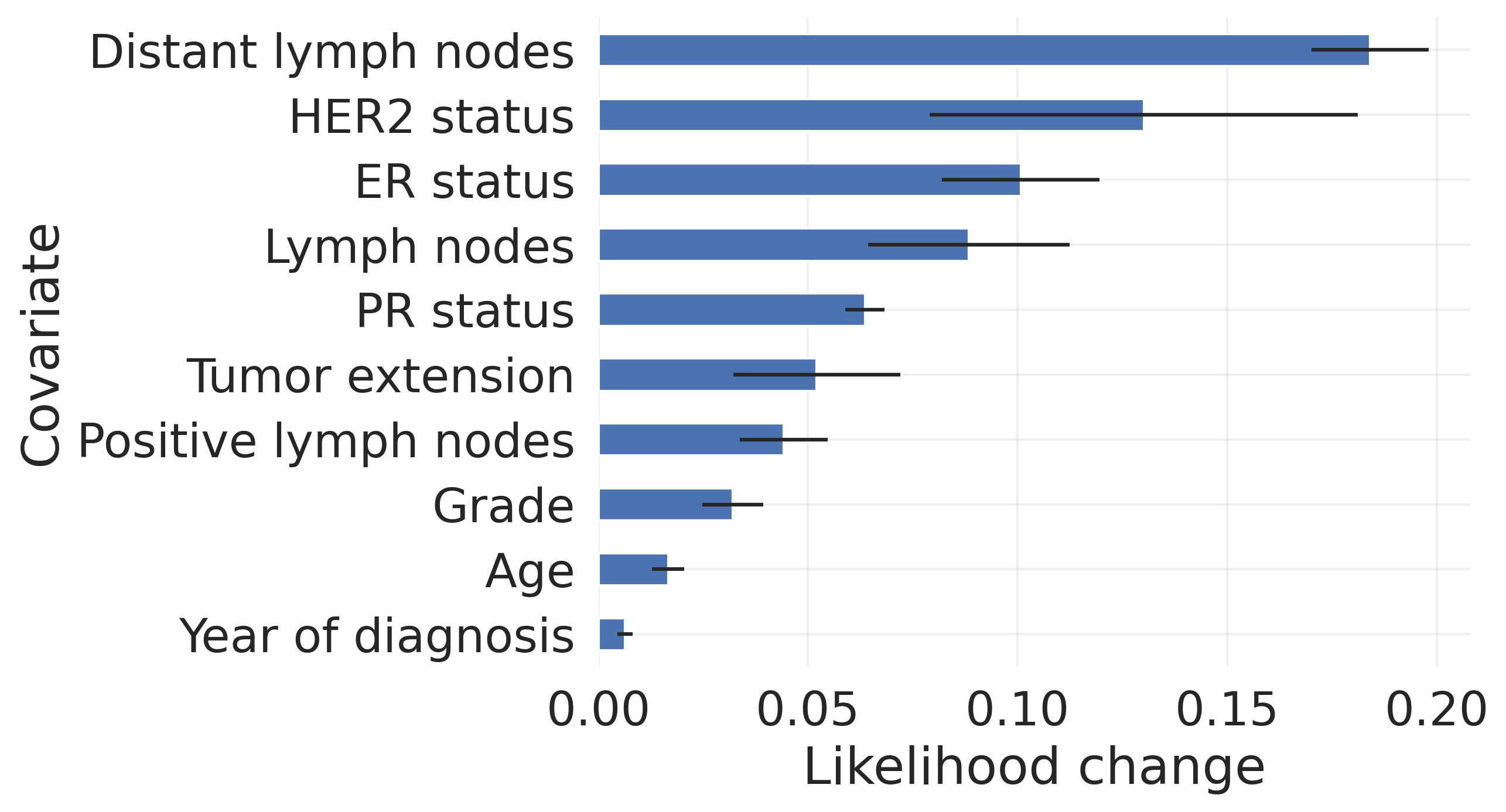}
    \caption{Causal Survival Clustering - Change in log-likelihood given random permutation of a given covariate. }
    \label{fig:cancer:importance}
\end{figure*}

Using a permutation test, we identify the covariates that most impact the likelihood associated with the proposed CSC model. Figure~\ref{fig:cancer:importance} displays the 10 covariates which most impact the likelihood, and therefore are associated with the different treatment response subgroups. This proposed analysis provides a tool to validate the model's medical relevance by studying which covariates are associated with outcomes.

\subsection{Baselines' identified subgroups}

\begin{figure*}[!htb]
    \centering
    \subfigure[Cox Mixture of Heterogeneous Effects]{\includegraphics[width=0.3\textwidth]{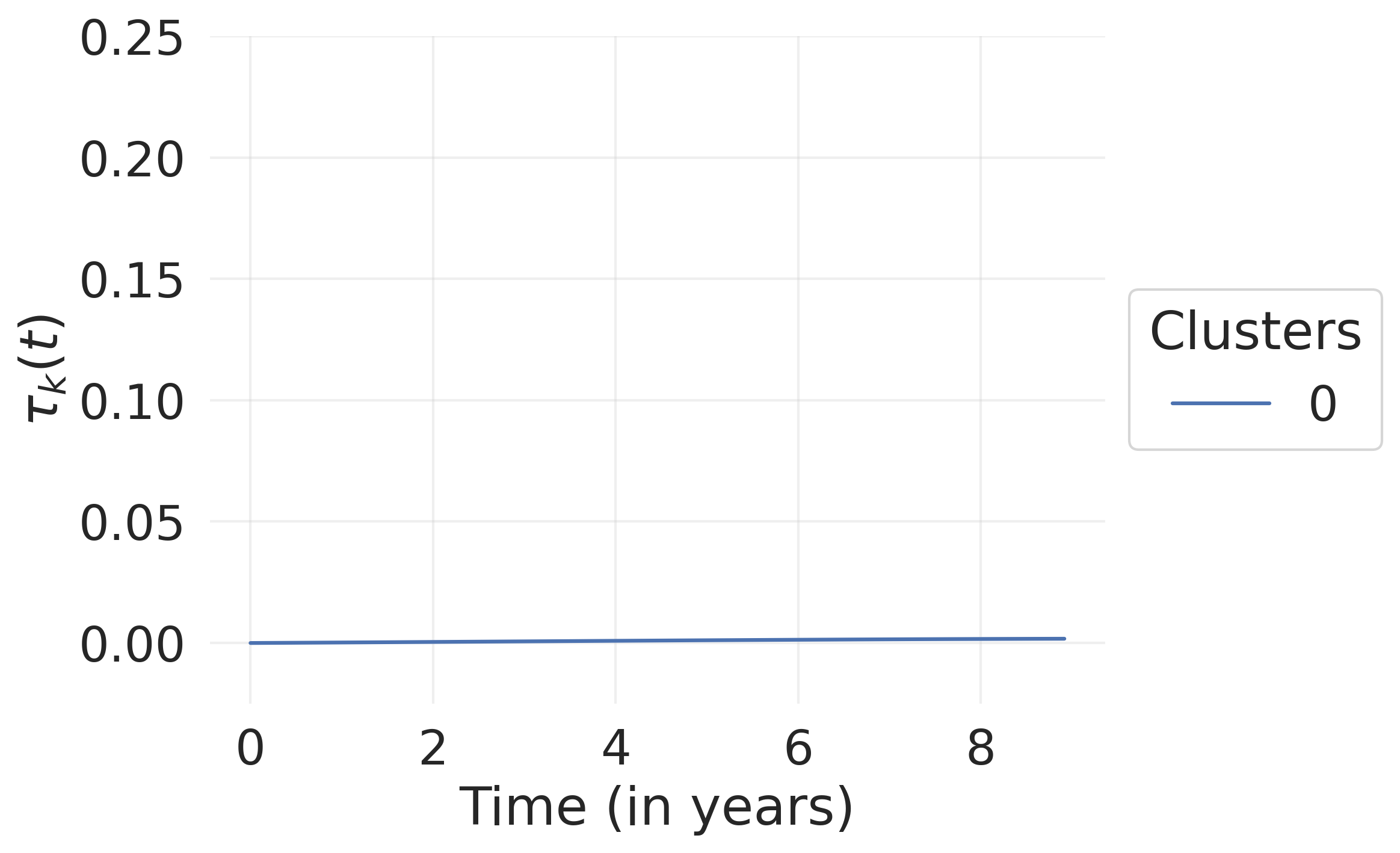}}
    \subfigure[KMeans]{\includegraphics[width=0.3\textwidth]{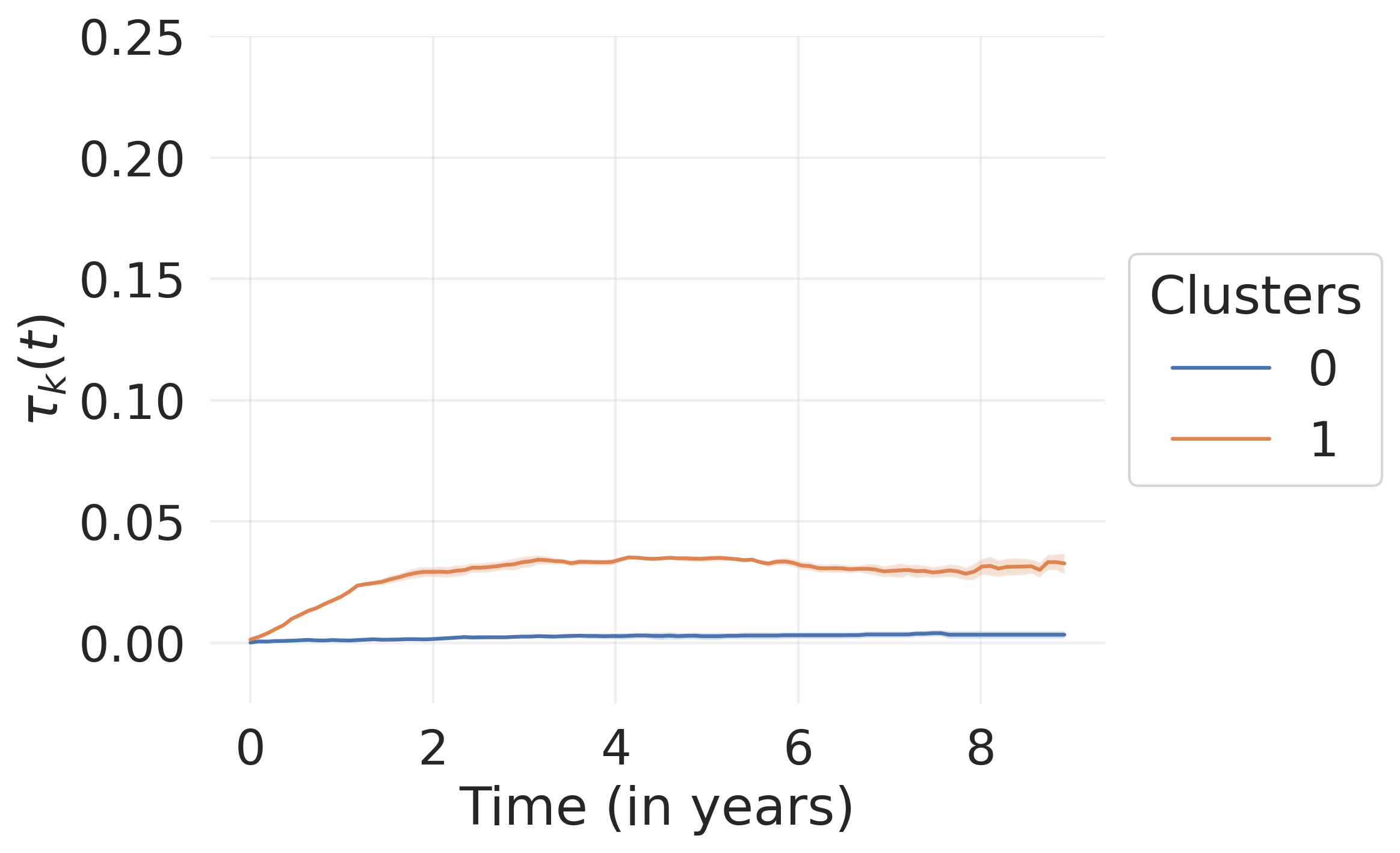}}
    \subfigure[Virtual Twins]{\includegraphics[width=0.3\textwidth]{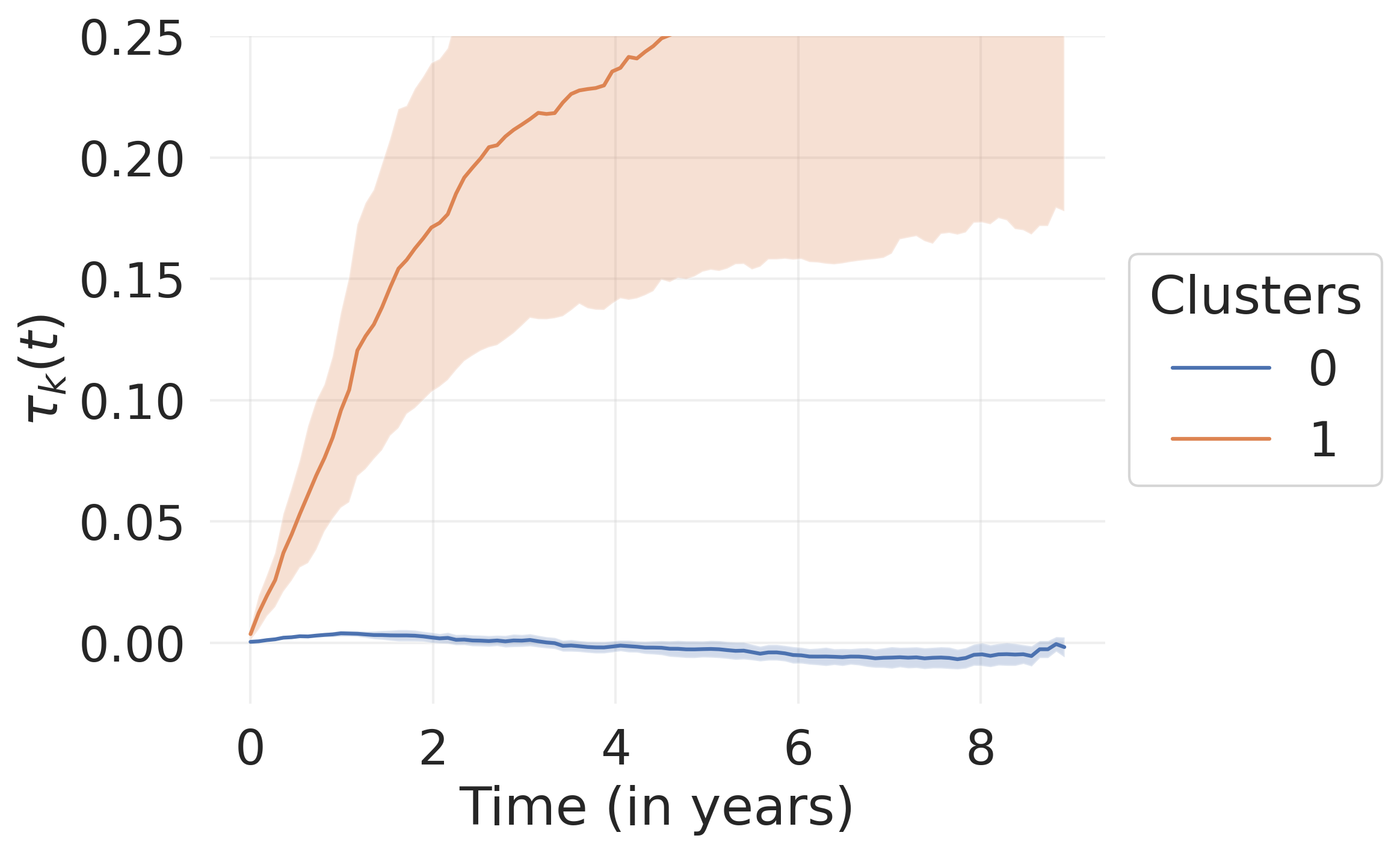}}
    \caption{Averaged treatment effect subgroups across 5-fold cross-validation observed in the \textsc{Seer} dataset with the shaded areas representing 95\% CI.}
    \label{ntc:fig:seer:other}
\end{figure*}
Our main analysis describes how to use the proposed methodology to study a medical application. In this section, we analyse the groups one would identify with the previously described baselines. Figure~\ref{ntc:fig:seer:other} displays the identified treatment effect subgroups using all considered baselines. For CMHE, the number of subgroups $K$ is selected through hyperparameter tuning, leading to $K = 1$ in every fold. As previously mentioned, our proposed methodology presents two strengths that explain the difference in the identified subgroups of treatment effects compared to CMHE. First, the survival distribution under treatment is not constrained by the one under the control regime, resulting in more flexible, non-proportional distributions. CMHE's parametrisation, which characterises treatment as a linear shift in the log hazard, results in a proportionality assumption between treated and untreated distributions. Second, CMHE does not account for treatment non-randomisation in its average treatment effect estimation, whereas our use of inverse propensity weighting corrects for any observed ones. While both methodologies identify a group with limited response, our proposed methodology distinguishes a second group with improved treatment response.

Additionally, the KMeans approach identifies clusters that resemble those found with CSC, though the detected positive response is comparatively smaller. In contrast, the virtual twins method identifies a subgroup that exhibits a larger treatment effect, while the majority shows a negative response to treatment. Importantly, both methods broadly corroborate our findings. However, CSC presents a key strength by jointly learning group membership and treatment effects, ensuring that the covariates defining each subgroup are directly associated with the outcome -- a property that two-step approaches cannot guarantee. 

\end{document}